\documentclass[12pt]{article}
\pdfoutput=1

\usepackage{amsmath,amssymb,amscd}
\usepackage{listings}
\usepackage{caption}
\usepackage{dsfont}
\usepackage{slashed}
\usepackage[dvipsnames]{xcolor}

\usepackage[pdftex]{graphicx}
\usepackage{epstopdf}
\usepackage{subfigure}
\usepackage{epsfig}
\usepackage{listings}
\usepackage{caption}
\usepackage{cite}
\usepackage{verbatim}
\usepackage{multicol}
\usepackage[colorlinks, citecolor=black, linkcolor=black]{hyperref}
\usepackage{feynmp}
\usepackage{multirow}
\usepackage[titletoc]{appendix}

\usepackage{array}
\newcolumntype{M}[1]{>{\centering\arraybackslash}m{#1}}
\newcolumntype{N}{@{}m{0pt}@{}}

\allowdisplaybreaks

\newcommand\T{\rule{0pt}{2.6ex}}       
\newcommand\B{\rule[-1.2ex]{0pt}{0pt}} 

\newcommand{\bea}{\begin{align}}
\newcommand{\eea}{\end{align}}
\newcommand{\beq}{\begin{equation}}
\newcommand{\eeq}{\end{equation}}
\newcommand{\nbea}{\begin{align*}}
\newcommand{\neea}{\end{align*}}
\newcommand{\nbeq}{\begin{equation*}}
\newcommand{\neeq}{\end{equation*}}
\newcommand{\bear}{\begin{eqnarray}}  
\newcommand{\eear}{\end{eqnarray}}  

 \newcommand{\twomatrix}[1]{\left(\begin{array}{cc} #1 \end{array}\right) }
 \newcommand{\column}[1]{\left(\begin{array}{c} #1 \end{array}\right) }
 
 \newcommand{\identity}{\mathds{1}}

\newcommand{\Dfbu}{\mathord{\buildrel{\lower3pt\hbox{$\scriptscriptstyle{\leftrightarrow \tiny{ \ \ \ } }$}}\over {D^{\mu}}}} 
\newcommand{\Dfbd}{\mathord{\buildrel{\lower3pt\hbox{$\scriptscriptstyle\leftrightarrow$}}\over {D}_{\mu}}} 

\numberwithin{equation}{section}

\begin{document}

\begin{titlepage}

\pagestyle{empty}

\baselineskip=21pt
\rightline{\footnotesize MCTP-17-09, KCL-PH-TH/2017-30, Cavendish-HEP-17/10, DAMTP-2017-25}
\vskip 0.6in

\begin{center}

{\large {\bf Extending the Universal One-Loop Effective Action: Heavy-Light Coefficients}}

\vskip 0.4in

 {\bf Sebastian~A.~R.~Ellis}$^{1}$,
 {\bf J\'er\'emie~Quevillon}$^{2}$,
{\bf Tevong~You}$^{3}$
and {\bf Zhengkang~Zhang}$^{1}$

\vskip 0.3in

{\small {\it
$^1${Michigan Center for Theoretical Physics (MCTP), \\ Department of Physics, University of Michigan, \\Ann Arbor, MI 48109, USA}\\
\vspace{0.25cm}
$^2${Theoretical Particle Physics and Cosmology Group, Physics Department, \\
King's College London, London WC2R 2LS, UK}\\
\vspace{0.25cm}
$^3${DAMTP, University of Cambridge, Wilberforce Road, Cambridge, CB3 0WA, UK; \\
Cavendish Laboratory, University of Cambridge, J.J. Thomson Avenue, \\ 
\vspace{-0.25cm}
Cambridge, CB3 0HE, UK}
}}

\vskip 0.4in

{\bf Abstract}

\end{center}

\baselineskip=18pt \noindent



 The Universal One-Loop Effective Action (UOLEA) is a general expression for the effective action obtained by evaluating in a model-independent way the one-loop expansion of a functional path integral. It can be used to match UV theories to their low-energy EFTs more efficiently by avoiding redundant steps in the application of functional methods, simplifying the process of obtaining Wilson coefficients of operators up to dimension six. In addition to loops involving only heavy fields, matching may require the inclusion of loops containing both heavy and light particles. Here we use the recently-developed covariant diagram technique to extend the UOLEA to include heavy-light terms which retain the same universal structure as the previously-derived heavy-only terms.
As an example of its application, we integrate out a heavy singlet scalar with a linear coupling to a light doublet Higgs. The extension presented here is a first step towards completing the UOLEA to incorporate all possible structures encountered in a covariant derivative expansion of the one-loop path integral. 



\vskip 0.5in

{\small \leftline{June 2017}}

\end{titlepage}

\newpage



\section{Introduction}
\label{sec:introduction}

Effective Field Theories (EFTs) have always played a prominent role in particle phenomenology~\cite{weinberghistory, georgihistory}. From early applications in the Heisenberg-Euler action of QED~\cite{heisenbergeuler}, the Fermi theory of weak interactions~\cite{fermitheory}, or the chiral Lagrangian of pions~\cite{chiralpion}, the framework has since been used to describe neutrino~\cite{weinbergoperator}, nuclear~\cite{nuclearEFT}, flavour~\cite{flavourEFT}, electroweak~\cite{electroweakEFT} and Higgs physics~\cite{higgsEFT}, to name a few examples. More recent developments include EFTs for dark matter~\cite{darkmatterEFT}, inflation~\cite{inflationEFT} and cosmology~\cite{cosmologyEFT}, as well as the Standard Model EFT (SM EFT)~\cite{higgsEFT, SMEFT}.

Obtaining a low-energy EFT by integrating out heavy degrees of freedom from an ultraviolet (UV) theory is typically performed using Feynman diagram methods. This involves calculating correlation functions among the light fields in the UV theory, expanding them in inverse powers of a heavy mass scale, then extracting the relevant parts for the Wilson coefficients by comparing with the same correlation functions computed in the EFT. Almost all instructions on matching in particle phenomenology follow this procedure~\footnote{See for example Ref.~\cite{electroweakEFT} for a pedagogical introduction to matching in EFTs.}.

This traditional Feynman diagram approach, albeit familiar and well-developed, is a rather roundabout route from $\mathcal{L}_\text{UV}$ to $\mathcal{L}_\text{EFT}$. In contrast, there are more elegant alternative methods for obtaining Wilson coefficients, which {\it avoid the need for computing correlation functions}. They are based on direct evaluation of the functional path integral, and various techniques exist for doing so up to one-loop level~\cite{Gaillard, Cheyette, mixedfunctional1, mixedfunctional2, HKMN, HLM, UOLEA, BGP, HLM2, UOLEAHL, Valencia, covariantdiagrams}~\footnote{Some recent functional matching calculations can be found in Refs.~\cite{matchingexamples}.}. In practice, functional methods have typically been overlooked in favour of the more traditional Feynman diagram approach, where many tools have been developed to ease otherwise complicated calculations. Nevertheless, Feynman diagram matching remains cumbersome for systematic derivations of a full set of Wilson coefficients, while recent developments in functional methods, which we summarise below, have led to a more straightforward way of dealing with the one-loop path integral. Moreover, the result of evaluating this path integral has a {\it universal form} that is independent of the method used to obtain it, suggesting a redundancy in historically repeating this evaluation with various different functional techniques. The logical step is then to eliminate such unnecessary calculations by doing them {\it once-and-for-all}. The expression obtained in this way is the so-called Universal One-Loop Effective Action (UOLEA)~\cite{HLM, UOLEA}. It allows one to bypass the need for either Feynman diagram or functional methods entirely when deriving Wilson coefficients for operators up to dimension six.

These developments began with a review of Gaillard~\cite{Gaillard} and Cheyette's~\cite{Cheyette} functional Covariant Derivative Expansion (CDE) method by Henning, Lu and Murayama (HLM)~\cite{HLM}. In particular, they noticed that for a simplified case where the multiplet of heavy fields are assumed degenerate, the resulting one-loop effective action could be evaluated with loop integrals factored out independently of the UV-specific parts. The expressions for the various combinations of the loop integrals could then be pre-evaluated and encapsulated into ``universal'' coefficients associated to terms involving the trace of matrices of light fields and commutators of covariant derivatives. Unfortunately, these results only applied to the special case of degenerate multiplet masses, meaning that for more general UV models one would have had to return to evaluating the path integral, or use Feynman diagrams, to obtain the one-loop effective action. However, in Ref.~\cite{UOLEA} some of us (JQ and TY, together with Drozd and J. Ellis) showed that the universality of the matrix terms and their associated coefficients also holds in the non-degenerate case, and derived the general UOLEA relevant for all operator structures up to dimension six, without any assumptions on the mass spectrum.

Another potential limitation was pointed out by Ref.~\cite{AKS}, following arguments from Ref.~\cite{bilenkysantamaria}, that functional methods did not appear to account for one-loop matching involving both heavy and light particles in the loop. This was addressed by us~\cite{UOLEAHL} and others~\cite{BGP, HLM2,Valencia}, each demonstrating different procedures for treating mixed heavy-light matching in the path integral approach~\footnote{Other more complicated functional methods for heavy-light matching had also been proposed in the past~\cite{mixedfunctional1, mixedfunctional2}.}. We emphasised in particular that our method also allowed for the computation of universal terms~\cite{UOLEAHL}, that could in principle be added to the original UOLEA. 

In this paper, we explicitly include a complete set of such universal heavy-light terms which retain the same structure as the previously-derived heavy-only terms. The results presented here serve as a systematic extension of the heavy-only UOLEA of Ref.~\cite{UOLEA}, thus settling definitively the question of whether the applicability of functional methods and their simplification due to universality could be extended to the heavy-light case.

Explicitly achieving such an extension requires computing a large number of terms in a CDE of the path integral, which would have been impractical (if not impossible) within previously proposed frameworks. However, a diagrammatic reformulation of functional matching recently developed by one of us (ZZ)~\cite{covariantdiagrams} greatly simplifies the task. It is now feasible to systematically extend the UOLEA and derive all of its associated universal coefficients, even by hand. In short, the idea of~\cite{covariantdiagrams} is to represent the CDE series as a sum of ``covariant diagrams'' which help organise the expansion in a systematic way. The spirit is similar to using traditional Feynman diagrams to keep track of expansions of correlation functions, but the key difference is that covariant diagrams evaluate directly to gauge-invariant operators in the EFT (as opposed to correlation functions). Moreover, the same universal structure of the UOLEA for both heavy-only and heavy-light terms is now put on firmer theoretical ground. In fact, the key step of expansion by regions (as introduced by~\cite{Valencia}) in the derivation of covariant diagrams makes it clear that heavy-light terms follow from heavy-only terms with simple substitutions, which we will show.

The paper is organised as follows. Section~\ref{sec:UOLEA} summarises for convenience both the previous heavy-only UOLEA and its current extension in order to clarify the relation of this extension to our previous (and future) work. Section~\ref{sec:universalcoefficients} lists the universal coefficients and describes the derivation and cross-checks we have made in their calculation. Section~\ref{sec:example} gives an application to integrating out a real scalar singlet with the heavy-light one-loop contributions computed here for the first time. Finally, we conclude with our perspective and outlook in Section~\ref{sec:conclusion}. The master integrals involved in the universal coefficients are discussed in more detail in Appendix~\ref{Integrals.APP}. Explicit expressions of the universal coefficients in the special case of degenerate heavy particles are collected in Appendix~\ref{app:uc}, while complete non-degenerate expressions can be found in a \texttt{Mathematica} notebook in the arXiv submission.

\section{The Universal One-Loop Effective Action}
\label{sec:UOLEA}

\subsection{Heavy-only UOLEA} 

Consider a UV Lagrangian involving a multiplet of heavy fields $\Phi$ coupled to light fields $\phi$, which for bosons may be arranged into the form
\begin{equation}
\mathcal{L}_\text{UV}[\phi,\Phi] = \mathcal{L}[\phi] + (\Phi^\dagger F[\phi] + \text{h.c.}) + \Phi^\dagger(P^2 - M^2 - U_H[\phi])\Phi + \mathcal{O}(\Phi^3) \, ,
\label{eq:lagrangianUV}
\end{equation}
where $P_\mu \equiv iD_\mu$ and $M$ is the (diagonalised) mass matrix for the multiplet $\Phi$. The linear coupling to light fields is parametrised by $F[\phi]$. It gives tree-level contributions to the Wilson coefficients of the effective action when substituting the equations of motion. The quadratic coupling to light fields is specified by the matrix $U_H[\phi]$, and we assume for now no additional dependence in this matrix on $P_\mu$ or gamma matrices. While the UV Lagrangian for fermionic fields is of a different form to Eq.~\ref{eq:lagrangianUV}, this general strategy of functional matching applies equally to fermions since the UV dependence is encapsulated in the same form as the bosonic case at the level of the logarithm expression (\ref{eq:logL}), derived below~\footnote{For details on how to rearrange the fermionic Lagrangian into this form, see for example Appendix A1 of \cite{HLM} and Appendix E of \cite{UOLEA}. }.

In the functional method for integrating out heavy fields, keeping only the light fields in an effective action, one evaluates the path integral for the quadratic part of the action in $\Phi$ expanded around its minimum (or background field value)~\footnote{We assume here a real scalar but this applies equally to complex or ghost scalar fields, as well as vectors and fermions~\cite{HLM, UOLEA, covariantdiagrams}.},   
\begin{align*}
e^{iS_\text{eff}[\phi]} &= \int [D\Phi] e^{iS[\phi,\Phi]} \\
&= \int [D\eta] e^{i\left(S[\phi,\Phi_c] + \frac{1}{2}\left.\frac{\delta^2 S}{\delta \Phi^2}\right|_{\Phi=\Phi_c}\eta^2 + \mathcal{O}(\eta^3)\right)} \\
&\approx e^{iS[\phi,\Phi_c]}\left[\text{det}\left(\left.-\frac{\delta^2 S}{\delta\Phi^2}\right|_{\Phi=\Phi_c}\right)\right]^{-\frac{1}{2}}	\\
&= e^{iS[\phi,\Phi_c] - \frac{1}{2}\text{Tr ln}\left(-\left.\frac{\delta^2 S}{\delta \Phi}\right|_{\Phi=\Phi_c}\right)} \, ,
\end{align*} 
where $\Phi_c$ is defined as $\left.\frac{\delta S}{\delta \Phi}\right|_{\Phi = \Phi_c} = 0$. This standard procedure relies on the Gaussian form of the functional integral for the quadratic term, and an identity for the determinant written in terms of a logarithmic operator in the action. The one-loop effective Lagrangian corresponding to $S_\text{eff}[\phi]$ is then 
\begin{equation}
\mathcal{L}_\text{EFT}^\text{1-loop}[\phi] = i c_s \int \frac{d^d q}{(2\pi)^d} \text{tr } \text{ln}\left(-P^2 + M^2 + U_H \right) \, ,
\label{eq:logL}
\end{equation}
which applies equally for heavy-only contributions from bosons and fermions, if $U_H$ is block-diagonal with respect to bosons vs.\ fermions \footnote{Additional care is required to take into account more complex structures that can potentially arise, as we discuss in Section~\ref{sec:completeUOLEA}. }. 
The lower-case trace is over all internal indices and the spacetime trace gives the momentum integral, expressed here in $d$ dimensions. The numerical pre-factor $c_s$ depends on the type of boson or fermion being integrated out~\cite{HLM}. 

The expansion of the logarithm in terms of a series of local operators suppressed by a heavy mass scale can be performed by a variety of techniques. As mentioned in the introduction, recent developments have led to a simple and systematic way of directly writing down the terms in this expansion using covariant diagrams. Regardless of the method used to evaluate the logarithm expansion, it can be done once-and-for-all, and the result is the same and universal in the sense that the final expression is independent of the details of the UV Lagrangian, which remain encapsulated in the $U_H$ matrix of light fields, covariant derivative $P_\mu$, and mass matrix $M$. This heavy-only universal one-loop effective action (UOLEA) can then be written as~\cite{UOLEA}
\begin{align}
\mathcal{L}_\text{UOLEA} =& -i c_s \,\mathrm{tr}\, \Bigl\{ 
f_2^i\, {U_H}_{ii} 
+f_3^i\, G^{\prime\mu\nu}_i G^\prime_{\mu\nu,i} 
+f_4^{ij}\, {U_H}_{ij} {U_H}_{ji} \nonumber\\ 
&\qquad\quad
+f_5^i\, [P^\mu, G^\prime_{\mu\nu,i}] [P_\rho, G^{\prime\rho\nu}_i]
+f_6^i\, G^{\prime\mu}_{\;\;\,\nu,i} G^{\prime\nu}_{\;\;\,\rho,i} G^{\prime\rho}_{\;\;\,\mu,i} \nonumber\\
&\qquad\quad
+f_7^{ij}\, [P^\mu, {U_H}_{ij}] [P_\mu, {U_H}_{ji}]
+f_8^{ijk}\, {U_H}_{ij} {U_H}_{jk} {U_H}_{ki}
+f_9^i\, {U_H}_{ii} G^{\prime\mu\nu}_i G^\prime_{\mu\nu,i} \nonumber\\ 
&\qquad\quad
+f_{10}^{ijkl}\, {U_H}_{ij} {U_H}_{jk} {U_H}_{kl} {U_H}_{li}
+f_{11}^{ijk}\, {U_H}_{ij} [P^\mu, {U_H}_{jk}] [P_\mu, {U_H}_{ki}] \nonumber\\ 
&\qquad\quad
+f_{12}^{ij}\, \bigl[P^\mu, [P_\mu, {U_H}_{ij}]\bigr] \bigl[P^\nu, [P_\nu, {U_H}_{ji}]\bigr]
+f_{13}^{ij}\, {U_H}_{ij} {U_H}_{ji} G^{\prime\mu\nu}_i G^\prime_{\mu\nu,i} \nonumber\\ 
&\qquad\quad
+f_{14}^{ij}\, [P^\mu, {U_H}_{ij}] [P^\nu, {U_H}_{ji}] G^\prime_{\nu\mu, i}
\nonumber\\
&\qquad\quad
+f_{15}^{ij}\, \bigl( {U_H}_{ij} [P^\mu, {U_H}_{ji}] -[P^\mu, {U_H}_{ij}] {U_H}_{ji} \bigr) [P^\nu, G^\prime_{\nu\mu,i}] \nonumber\\ 
&\qquad\quad
+f_{16}^{ijklm}\, {U_H}_{ij} {U_H}_{jk} {U_H}_{kl} {U_H}_{lm} {U_H}_{mi} \nonumber\\ 
&\qquad\quad
+f_{17}^{ijkl}\, {U_H}_{ij} {U_H}_{jk} [P^\mu, {U_H}_{kl}] [P_\mu, {U_H}_{li}]
+f_{18}^{ijkl}\, {U_H}_{ij} [P^\mu, {U_H}_{jk}] {U_H}_{kl} [P_\mu, {U_H}_{li}] \nonumber\\ 
&\qquad\quad
+f_{19}^{ijklmn}\, {U_H}_{ij} {U_H}_{jk} {U_H}_{kl} {U_H}_{lm} {U_H}_{mn} {U_H}_{ni}
\Bigr\} .
\label{eq:heavyonlyUOLEA}
\end{align}
The universal coefficients $f_N$ contain combinations of master integrals, which are defined and discussed in more detail in Appendix \ref{Integrals.APP}, and are listed in Tables~\ref{tab:PonlyandU2P2andU1P4}, \ref{tab:U3P2}, \ref{tab:U2P4}, \ref{tab:U4P2}, \ref{tab:Uonly}. Higher-order terms in the expansion may be computed but this expression is sufficient for all operators up to dimension 6. To obtain the Wilson coefficients of operators in a specific EFT, one then substitutes in to Eq.~\ref{eq:heavyonlyUOLEA} the particular $U_H$ matrix of light fields, covariant derivatives $P_\mu$ and mass matrix $M$ for a particular UV model. The result can then be brought into a non-redundant EFT basis if desired.

\subsection{Heavy-light UOLEA}   

To perform one-loop matching including cases where both heavy and light fields enter in the loop, we also expand the quantum fluctuations of the light fields around their background. The covariant derivative $P_\mu$, mass matrix $M$, and quadratic field matrix $U$ are extended accordingly. Through appropriate functional manipulations, detailed for example in Ref.~\cite{covariantdiagrams}, the general form of the Lagrangian can then be most conveniently written in the form
\begin{equation}
\mathcal{L}_\text{EFT}^\text{1-loop}[\phi] = i c_s \int \frac{d^d q}{(2\pi)^d} \text{tr } \text{ln} \left. \left(\Delta_H  - U_{HL}\Delta_L^{-1}U_{LH}  \right) \right|_\text{hard} \, ,
\label{eq:extendedonelooplagrangian}
\end{equation}
where
\begin{equation}
\Delta \equiv -P^2 + M^2 + U \, ,
\label{eq:Delta}
\end{equation}
and $P$, $M$ and $U$ are for the heavy or light fields depending on the subscript of $\Delta$. Only the ``hard'' part of the integrals are kept when using the integration by regions method to evaluate the integrals~\footnote{The separation of the integrals into hard and soft parts in the integration by regions method is well described in Refs.~\cite{Valencia, covariantdiagrams}.}. In addition to the heavy-only $\Delta_H$ part, we now also have terms corresponding to the heavy-light loops contributing to the expansion.

The universal form of this expansion has the same structure as the heavy-only UOLEA of Eq.~\ref{eq:heavyonlyUOLEA}, with more possibilities corresponding to different insertions of heavy, light, and heavy-light $U$ matrices. We therefore number the resulting universal coefficients according to their corresponding structure in Eq.~\ref{eq:heavyonlyUOLEA} with a different alphabetical letter for each variation on that structure. 

Our extended UOLEA is then a sum of universal terms and their associated coefficients,
\begin{equation}
{\cal L}^{\text{eff}}_{\text{1-loop}}[\phi] \supset -i c_s \sum_N f_N \, \mathrm{tr}\;\mathbb{O}_N\, ,
\end{equation}
where the universal coefficients $f_N$ correspond to the universal operator structures $\mathbb{O}_N$ and are labelled by $N=2, 3, 4, 4A,\ldots$, and contain combinations of the master integrals defined and discussed in Appendix \ref{Integrals.APP}. The universal operators $\mathbb{O}_N$ are combinations of $P, U_H, U_L, U_{HL}$ and $U_{LH}$ matrices. These universal coefficients and structures are listed in Tables~\ref{tab:PonlyandU2P2andU1P4}, \ref{tab:U3P2}, \ref{tab:U2P4}, \ref{tab:U4P2}, \ref{tab:Uonly} and in Section~\ref{sec:universalcoefficients} we describe in detail the derivation of our results.

\subsection{Towards a Complete UOLEA}   
\label{sec:completeUOLEA}
  
In Ref.~\cite{UOLEAHL} we also identified further structures that occur in the quadratic term of the action for specific cases, for example when integrating out heavy particles coupling with {\it ``open'' covariant derivatives}, i.e. covariant derivatives which do not appear in commutators. In this case the most general form of Eq.~\ref{eq:Delta} involves additional terms with $Z$ matrices of the form 
\begin{equation}
\Delta \equiv -P^2 + M^2 + U[\Phi,\phi] + P_\mu Z^\mu[\Phi, \phi] + {Z^\dagger}^\mu[\Phi,\phi]P_\mu + \text{...} \, .
\end{equation}
In addition, if the UV theory contains both bosons and fermions, and the $U$, $Z$ matrices are not block-diagonal with respect to bosons vs.\ fermions, it is preferable to take an alternative route in functional matching, treating bosons and fermions differently as in~\cite{covariantdiagrams}. In this case, additional universal terms can arise.

These additional structures were previously neglected in Refs.~\cite{HLM, UOLEA} and are not yet incorporated in the heavy-light UOLEA presented in this paper, but will be included in future work~\cite{workinprogress}. As shown in~\cite{covariantdiagrams}, covariant diagrams are capable of dealing with them. Therefore, with the necessary techniques at hand, we have a clear path toward a complete UOLEA that includes all possible structures encountered in a covariant derivative expansion of the one-loop path integral.

\section{Universal Coefficients}    
\label{sec:universalcoefficients}

\begin{table}
{\small
\begin{center}
\begin{tabular}
{ | c | c |   } 
\hline
\multicolumn{2}{|c|}{ \T\B $P$-only terms }  \\
\hline
\T\B $f_3^{i} = 2\,\mathcal{I}[q^4]_{i}^{4}$ & ${G^\prime}^{\mu\nu}_i{G^\prime_{\mu\nu}}_i $ \\
\hline
\T\B $f_5^{i} = 16\,\mathcal{I}[q^6]_{i}^{6}$ & $[P^\mu,{G^\prime_{\mu\nu}}_i][P_\rho,{G^\prime}^{\rho\nu}_i] $ \\
\hline
\T\B $f_6^{i} = (32/3)\,\mathcal{I}[q^6]_{i}^{6}$ & ${{{G^\prime}^\mu}_\nu}_i{{{G^\prime}^\nu}_\rho}_i{{{G^\prime}^\rho}_\mu}_i $ \\
\hline
\end{tabular}
\begin{tabular}
{ | c | c |   } 
\hline
\multicolumn{2}{|c|}{ \T\B $\mathcal{O}(U_H^2 P^2)$ terms }  \\
\hline
\T\B $f_7^{ij} = \mathcal{I}[q^2]_{ij}^{22}$ & $[P^\mu, {U_H}_{ij}][P_\mu,{U_H}_{ji}]$ \\
\hline
\hline
\multicolumn{2}{|c|}{ \T\B $\mathcal{O}(U_{HL}^1U_{LH}^1 P^2)$ terms }  \\
\hline
\T\B $f_{7A}^{i} = 2\,\mathcal{I}[q^2]_{i0}^{22} $ & $[P^{\mu},{U_{HL}}_{ii^\prime}][P_\mu,{U_{LH}}_{i^\prime i}]$ \\
\hline
\end{tabular}
\end{center}
\begin{center}
\begin{tabular}
{ | c | c |   } 
\hline
\multicolumn{2}{|c|}{ \T\B $\mathcal{O}(U_H^1 P^4)$ terms }  \\
\hline
\T\B $f_9^{i} = 8\,\mathcal{I}[q^4]_{i}^{5}$ & ${U_H}_{ii} {G^\prime}^{\mu\nu}_i{G^\prime_{\mu\nu}}_i $ \\
\hline
\end{tabular}
\end{center}
}
\caption{\it Left: Universal coefficients $f_3^{i}$, $f_5^{i}$, and $f_6^{i}$ for operators involving only $P$. There are 3 heavy-heavy terms. Right: Universal coefficients $f_7^{ij}$ and $f_{7A}^{i}$ for operators at $\mathcal{O}(U^2 P^2)$. There are 2 terms, 1 heavy-heavy and 1 heavy-light. Centre: Universal coefficient $f_9^{i}$ for operators at $\mathcal{O}(U^1 P^4)$. There is only 1 heavy-heavy term.}
\label{tab:PonlyandU2P2andU1P4}
\end{table}

\begin{table}
{\small
\begin{center}
\begin{tabular}
{ | c | c |  } 
\hline
\multicolumn{2}{|c|}{ \T\B $\mathcal{O}(U_H^3 P^2)$ terms }  \\
\hline
\T\B $f_{11}^{ijk} = 2\left(\mathcal{I}[q^2]^{122}_{ijk} + \mathcal{I}[q^2]^{212}_{ijk}\right) $ & ${U_H}_{ij}[P^\mu,{U_H}_{jk}][P_\mu, {U_H}_{ki}]$ \\
\hline
\hline
\multicolumn{2}{|c|}{ \T\B  $\mathcal{O}(U_H^1 U_{HL}^1 U_{LH}^1 P^2)$ terms }  \\
\hline
\T\B $f_{11A}^{ij} = 2\left(\mathcal{I}[q^2]_{ij0}^{122} + \mathcal{I}[q^2]_{ij0}^{212}\right)$ & $ {U_H}_{ij}[P^\mu,{U_{HL}}_{ji^\prime}][P_\mu,{U_{LH}}_{i^\prime i}] $ \\
\T\B $f_{11B}^{ij} = 2\left(\mathcal{I}[q^2]_{ij0}^{221} + \mathcal{I}[q^2]_{ij0}^{122}\right)$  & $ {U_{LH}}_{i^\prime i}[P^\mu,{U_{H}}_{ij}][P_\mu,{U_{HL}}_{j i^\prime}] +  {U_{HL}}_{i i^\prime}[P^\mu,{U_{LH}}_{i^\prime j}][P_\mu,{U_{H}}_{j i}] $ \\
\hline
\hline
\multicolumn{2}{|c|}{ \T\B  $\mathcal{O}( U_L^1 U_{HL}^1 U_{LH}^1 P^2)$ terms }  \\
\hline
\T\B $f_{11C}^{i} = 4\,\mathcal{I}[q^2]_{i0}^{23}$  & $ {U_{L}}_{i^\prime j^\prime}[P^\mu,{U_{LH}}_{j^\prime i}][P_\mu,{U_{HL}}_{i i^\prime}] $ \\
\T\B $f_{11D}^{i} = 2\left(\mathcal{I}[q^2]_{i0}^{14} + \mathcal{I}[q^2]_{i0}^{23}\right)$  & $ {U_{HL}}_{i i^\prime}[P^\mu,{U_{L}}_{i^\prime j^\prime}][P_\mu,{U_{LH}}_{j^\prime i}] + {U_{LH}}_{i^\prime i}[P^\mu,{U_{HL}}_{i j^\prime}][P_\mu,{U_{L}}_{j^\prime i^\prime}] $ \\
\hline
\end{tabular}
\end{center}
}
\caption{\it Universal coefficients $f_{11}^{ijk}$, $f_{11A, B, C, D}^{ij}$ for operators at $\mathcal{O}(U^3 P^2)$. There are 5 terms: 1 heavy-heavy and 4 heavy-light. The heavy-light ones are sub-divided into two groups of 2 terms. }
\label{tab:U3P2}
\end{table}

\begin{table}
{\footnotesize 
\begin{center}
\begin{tabular}
{ | c | c |   } 
\hline
\multicolumn{2}{|c|}{ \T\B $\mathcal{O}(U_H^2 P^4)$ terms }  \\
\hline
\T\B $f_{12}^{ij} = 4\,\mathcal{I}[q^4]_{ij}^{33}$ & $[P^\mu,[P_\mu,{U_H}_{ij}]][P^\nu,[P_\nu,{U_H}_{ji}]]$ \\
\hline
\T\B $f_{13}^{ij} = 4\left(\mathcal{I}[q^4]_{ij}^{33} + 2\,\mathcal{I}[q^4]_{ij}^{42} + 2\,\mathcal{I}[q^4]_{ij}^{51}\right)$ & ${U_H}_{ij}{U_H}_{ji} {G^\prime}^{\mu\nu}_i{G^\prime_{\mu\nu}}_i$ \\
\hline
\T\B $f_{14}^{ij} = -8\,\mathcal{I}[q^4]_{ij}^{33}$ & $ [P^\mu,{U_H}_{ij}][P^\nu,{U_H}_{ji}]{G^\prime_{\nu\mu}}_i$ \\
\hline
\T\B $f_{15}^{ij} = 4\left(\mathcal{I}[q^4]_{ij}^{33} + \mathcal{I}[q^4]_{ij}^{42}\right)$ & $\left({U_H}_{ij}[P^\mu,{U_H}_{ji}] - [P^\mu,{U_H}_{ij}]{U_H}_{ji}\right) [P^\nu,{G^\prime_{\nu\mu}}_i] $ \\
\hline
\hline
\multicolumn{2}{|c|}{ \T\B $\mathcal{O}(U_{HL}^1 U_{LH}^1 P^4)$ terms }  \\
\hline
\T\B $f_{12A}^{i} = 8\,\mathcal{I}[q^4]_{i0}^{33}$ &  $[P^\mu,[P_\mu,{U_{HL}}_{ii^\prime}]][P^\nu,[P_\nu,{U_{LH}}_{i^\prime i}]]$ \\
\hline
\T\B  $f_{13A}^{i} = 4\left(\,\mathcal{I}[q^4]_{i0}^{33} + 2 \,\mathcal{I}[q^4]_{i0}^{42} + 2 \,\mathcal{I}[q^4]_{i0}^{51} \right) $ &  ${U_{HL}}_{ii^\prime}{U_{LH}}_{i^\prime i} {G^\prime}^{\mu\nu}_i{G^\prime_{\mu\nu}}_i$ \\
\T\B $f_{13B}^{i} = 4\left(\,\mathcal{I}[q^4]_{i0}^{33} + 2 \,\mathcal{I}[q^4]_{i0}^{24} + 2 \,\mathcal{I}[q^4]_{i0}^{15} \right) $ &  ${U_{LH}}_{i^\prime i}{U_{HL}}_{i i^\prime} {G^\prime}^{\mu\nu}_{i^\prime}{G^\prime_{\mu\nu}}_{i^\prime}$ \\
\hline
\T\B $f_{14A}^{i} =  - 8 \,\mathcal{I}[q^4]_{i0}^{33} $ & $ [P^\mu,{U_{HL}}_{i i^\prime}][P^\nu,{U_{LH}}_{i^\prime i}]{G^\prime_{\nu\mu}}_i +[P^\mu,{U_{LH}}_{i^\prime i}][P^\nu,{U_{HL}}_{i i^\prime}]{G^\prime_{\nu\mu}}_{i^\prime}$ \\
\hline
\T\B $f_{15A}^{i} = 4 \left( \,\mathcal{I}[q^4]_{i0}^{33} + \mathcal{I}[q^4]_{i0}^{42} \right)  $  &  $\left({U_{HL}}_{ii^\prime}[P^\mu,{U_{LH}}_{i^\prime i}] - [P^\mu,{U_{HL}}_{i i^\prime}]{U_{LH}}_{i^\prime i}\right) [P^\nu,{G^\prime_{\nu\mu}}_i] $ \\
$f_{15B}^{i} = 4 \left( \,\mathcal{I}[q^4]_{i0}^{33} + \mathcal{I}[q^4]_{i0}^{24} \right)  $ & $ \left({U_{LH}}_{i^\prime i}[P^\mu,{U_{HL}}_{i i^\prime}] - [P^\mu,{U_{LH}}_{i^\prime i}]{U_{HL}}_{i i^\prime}\right) [P^\nu,{G^\prime_{\nu\mu}}_{i^\prime}] $ \\
\hline
\end{tabular}
\end{center}
}
\caption{\it Universal coefficients $f_{12}^{ij}$, $f_{12A}^{ij}$, $f_{13}^{ij}$, $f_{13A,B}^{i}$, $f_{14}^{ij}$, $f_{14A}^{i}$, $f_{15}^{ij}$ and $f_{15A,B}^{ij}$ for operators at $\mathcal{O}(U^2 P^4)$. There are 10 terms, 4 heavy-heavy and 6 heavy-light.  }
\label{tab:U2P4}
\end{table}

\begin{table}
{\footnotesize 
\begin{center}
\begin{tabular}
{ | c | c |   } 
\hline
\multicolumn{2}{|c|}{ \T\B $\mathcal{O}(U_H^4 P^2)$ terms }  \\
\hline
\T\B $f_{17}^{ijkl} = 2\left(\mathcal{I}[q^2]_{ijkl}^{2112} + \mathcal{I}[q^2]_{ijkl}^{1212} + \mathcal{I}[q^2]_{ijkl}^{1122}\right) $ &  $ {U_{H}}_{ij}{U_{H}}_{jk}[P^\mu,{U_{H}}_{kl}][P_\mu,{U_{H}}_{li}] $ \\
\hline
\T\B $f_{18}^{ijkl} = \mathcal{I}[q^2]_{ijkl}^{2121} + \mathcal{I}[q^2]_{ijkl}^{2112} + \mathcal{I}[q^2]_{ijkl}^{1221} + \mathcal{I}[q^2]_{ijkl}^{1212} $ & ${U_{H}}_{ij}[P^\mu,{U_{H}}_{jk}]{U_{H}}_{kl}[P_\mu,{U_{H}}_{li}] $ \\
\hline
\hline
\multicolumn{2}{|c|}{ \T\B $\mathcal{O}(U_H^2 U_{HL}^1 U_{LH}^1 P^2)$ terms }  \\
\hline
\T\B \multirow{2}{*}{$f_{17A}^{ijk} = 2 \left( \mathcal{I}[q^2]_{ijk0}^{1122} + \mathcal{I}[q^2]_{ijk0}^{1221} + \mathcal{I}[q^2]_{ijk0}^{2121} \right) $ } & ${U_{H}}_{ij} {U_{HL}}_{j i^\prime} [P^\mu,{U_{LH}}_{i^\prime k}] [P_\mu, {U_{H}}_{ki}] $ \\
\T\B & $ \quad + {U_{LH}}_{i^\prime i} {U_{H}}_{ij} [P^\mu,{U_{H}}_{jk}] [P_\mu,{U_{HL}}_{k i^\prime}]$ \\
\hline
\T\B $f_{17B}^{ijk} = 2 \left(\mathcal{I}[q^2]_{ijk0}^{1122} + \mathcal{I}[q^2]_{ijk0}^{1212} + \mathcal{I}[q^2]_{ijk0}^{2112} \right) $ & ${U_{H}}_{ij} {U_{H}}_{jk} [P^\mu,{U_{HL}}_{k i^\prime}] [P_\mu,{U_{LH}}_{i^\prime i}]$ \\
\hline
\T\B $f_{17C}^{ijk}= 2 \left(\mathcal{I}[q^2]_{ijk0}^{1122} + \mathcal{I}[q^2]_{ijk0}^{1221} + \mathcal{I}[q^2]_{ijk0}^{2121} \right) $ & ${U_{HL}}_{i i^\prime} {U_{LH}}_{i^\prime j} [P^\mu,{U_{H}}_{jk}], [P_\mu,{U_{H}}_{ki}]$ \\
\hline
\T\B \multirow{2}{*}{$f_{18A}^{ijk} = 2\left( \mathcal{I}[q^2]_{ijk0}^{1221} + \mathcal{I}[q^2]_{ijk0}^{2121} + \mathcal{I}[q^2]_{ijk0}^{1212} + \mathcal{I}[q^2]_{ijk0}^{2112} \right) $ } & ${U_{H}}_{ij}[P^\mu,{U_{HL}}_{j i^\prime}]{U_{LH}}_{i^\prime k}[P_\mu,{U_{H}}_{k i}]$ \\
\T\B $ $ & $ \quad + {U_{H}}_{ij}[P^\mu,{U_{H}}_{jk}]{U_{HL}}_{k i^\prime}[P_\mu,{U_{LH}}_{i^\prime i}]$ \\
\hline
\hline
\multicolumn{2}{|c|}{ \T\B $\mathcal{O}(U_H^1 U_L^1 U_{HL}^1 U_{LH}^1 P^2)$ terms }  \\
\hline
\T\B \multirow{2}{*}{$f_{17D}^{ij} = 2 \left( 2 \,\mathcal{I}[q^2]_{ij0}^{123} + \mathcal{I}[q^2]_{ij0}^{222} \right) $ } & ${U_{HL}}_{i i^\prime} {U_{L}}_{i^\prime j^\prime} [P^\mu, {U_{LH}}_{j^\prime j}][P_\mu, {U_{H}}_{ji}]$ \\
\T\B  & $\quad + {U_{L}}_{i^\prime j^\prime} {U_{LH}}_{j^\prime i} [P^\mu, {U_{H}}_{ij}] [P_\mu, {U_{HL}}_{j i^\prime}]$ \\
\hline
\T\B \multirow{2}{*}{$f_{17E}^{ij} = 2 \left( \mathcal{I}[q^2]_{ij0}^{114} + \mathcal{I}[q^2]_{ij0}^{123} + \mathcal{I}[q^2]_{ij0}^{213} \right) $ } & ${U_{H}}_{ij} {U_{HL}}_{j i^\prime} [P^\mu, {U_{L}}_{i^\prime j^\prime}] [P_\mu, {U_{LH}}_{j^\prime i}]$ \\
\T\B  & $\quad + {U_{LH}}_{i^\prime i} {U_{H}}_{ij} [P^\mu, {U_{HL}}_{j j^\prime}] [P_\mu, {U_{L}}_{j^\prime i^\prime}]$ \\
\hline
\T\B $f_{18B}^{ij} = 2 \left( \mathcal{I}[q^2]_{ij0}^{123} + \mathcal{I}[q^2]_{ij0}^{222} + \mathcal{I}[q^2]_{ij0}^{114} + \mathcal{I}[q^2]_{ij0}^{213} \right) $ & ${U_{HL}}_{i i^\prime}[P^\mu,{U_{L}}_{i^\prime j^\prime}]{U_{LH}}_{j^\prime j}[P_\mu,{U_{H}}_{j i}]$ \\
\hline
\T\B $f_{18C}^{ij} = 4 \left( \mathcal{I}[q^2]_{ij0}^{123} + \mathcal{I}[q^2]_{ij0}^{213} \right) $ & ${U_{H}}_{ij}[P^\mu,{U_{HL}}_{j i^\prime}]{U_{L}}_{i^\prime j^\prime}[P_\mu,{U_{LH}}_{j^\prime i}]$ \\
\hline
\hline
\multicolumn{2}{|c|}{ \T\B $\mathcal{O}(U_L^2 U_{HL}^1 U_{LH}^1 P^2)$ terms }  \\
\hline
\T\B \multirow{2}{*}{$f_{17F}^{i} =  2 \left( 2\,\mathcal{I}[q^2]_{i0}^{15} + \mathcal{I}[q^2]_{i0}^{24} \right) $ } & ${U_{HL}}_{i i^\prime} {U_{L}}_{i^\prime j^\prime} [P^\mu, {U_{L}}_{j^\prime k^\prime}] [P_\mu, {U_{LH}}_{k^\prime i}]$ \\
\T\B & $\quad + {U_{L}}_{i^\prime j^\prime} {U_{LH}}_{j^\prime i} [P^\mu, {U_{HL}}_{i k^\prime}] [P_\mu, {U_{L}}_{k^\prime i^\prime}]$ \\
\hline
\T\B $f_{17G}^{i} =  2 \left( 2\,\mathcal{I}[q^2]_{i0}^{15} + \mathcal{I}[q^2]_{i0}^{24} \right) $ & ${U_{LH}}_{i^\prime i} {U_{HL}}_{i j^\prime} [P^\mu, {U_{L}}_{j^\prime k^\prime}] [P_\mu, {U_{L}}_{k^\prime i^\prime}]$ \\
\hline
\T\B $f_{17H}^{i} = 6\,\mathcal{I}[q^2]_{i0}^{24}  $ & ${U_{L}}_{i^\prime j^\prime} {U_{L}}_{j^\prime k^\prime} [P^\mu, {U_{LH}}_{k^\prime i}] [P_\mu, {U_{HL}}_{i i^\prime}]$ \\
\hline
\T\B \multirow{2}{*}{$f_{18D}^{i} =4 \left( \mathcal{I}[q^2]_{i0}^{15} + \mathcal{I}[q^2]_{i0}^{24} \right)  $ } & ${U_{HL}}_{i i^\prime}[P^\mu,{U_{L}}_{i^\prime j^\prime}]{U_{L}}_{j^\prime k^\prime}[P_\mu,{U_{LH}}_{k^\prime i}]$ \\
\T\B  & $ \quad + {U_{LH}}_{i^\prime i}[P^\mu,{U_{HL}}_{i j^\prime}]{U_{L}}_{j^\prime k^\prime}[P_\mu,{U_{L}}_{k^\prime i}]$ \\
\hline
\hline
\multicolumn{2}{|c|}{ \T\B $\mathcal{O}( U_{HL}^2 U_{LH}^2 P^2)$ terms }  \\
\hline
\T\B $f_{17I}^{ij} =  2 \left( \mathcal{I}[q^2]_{ij0}^{114} + \mathcal{I}[q^2]_{ij0}^{213} + \mathcal{I}[q^2]_{ij0}^{123} \right) $ & ${U_{HL}}_{i i^\prime} {U_{LH}}_{i^\prime j} [P^\mu, {U_{HL}}_{j j^\prime}] [P_\mu, {U_{LH}}_{j^\prime i}]$  \\
\hline
\T\B $f_{17J}^{ij} = 2 \left( \mathcal{I}[q^2]_{ij0}^{222} + 2 \,\mathcal{I}[q^2]_{ij0}^{123} \right) $ & ${U_{LH}}_{i^\prime i} {U_{HL}}_{i j^\prime} [P^\mu, {U_{LH}}_{j^\prime j}] [P_\mu, {U_{HL}}_{j i^\prime}]$ \\
\hline
\T\B \multirow{2}{*}{$f_{18E}^{ij} = \mathcal{I}[q^2]_{ij0}^{114} + \mathcal{I}[q^2]_{ij0}^{123} +\mathcal{I}[q^2]_{ij0}^{213} + \mathcal{I}[q^2]_{ij0}^{222} $ } & ${U_{HL}}_{i i^\prime}[P^\mu,{U_{LH}}_{i^\prime j}]{U_{HL}}_{j j^\prime}[P_\mu,{U_{LH}}_{j^\prime i}]$ \\
\T\B  & $ \quad + {U_{LH}}_{i^\prime i}[P^\mu,{U_{HL}}_{i j^\prime}]{U_{LH}}_{j^\prime j}[P_\mu,{U_{HL}}_{j i^\prime}]$ \\
\hline
\end{tabular}
\end{center}
}
\caption{\it Universal coefficients $f_{17}^{ijkl}, f_{18}^{ijkl}, f_{17A, B, C}^{ijk}, f_{17D, E}^{ij}, f_{17F, G, H}^{i}, f_{17I, J}^{ij}, f_{18A}^{ijk}, f_{18B, C}^{ij}, f_{18D}^{i}$ and $f_{18E}^{ij}$ for operators at $\mathcal{O}(U^4 P^2)$. There are 17 terms: 2 heavy-heavy ones, 3 groups of 4 heavy-light terms, and 1 group of 3 heavy-light terms.}
\label{tab:U4P2}
\end{table}

\begin{table}
{\footnotesize 
\begin{center}
\resizebox{\textwidth}{!}{
\begin{tabular}
{ | c | c  | c | c | } 
\hline
\multicolumn{2}{|c|}{ \T\B $\mathcal{O}(U)$ term } & \multicolumn{2}{c|}{ \T\B $\mathcal{O}(U^3)$ terms }   \\
\hline
\T\B $f_2^{i} = \mathcal{I}_{i}^{1}$ & ${U_{H}}_{ii}$ & $f_{8}^{ijk} = \frac{1}{3}\,\mathcal{I}_{ijk}^{111}$ & ${U_{H}}_{ij}{U_{H}}_{jk}{U_{H}}_{ki}$  \\
\hline
\multicolumn{2}{|c|}{ \T\B $\mathcal{O}(U^2)$ terms } & $f_{8A}^{ij} = \mathcal{I}_{ij0}^{111}$ & ${U_{H}}_{ij}{U_{HL}}_{ji^\prime}{U_{LH}}_{i^\prime i}$   \\
\hline
\T\B $f_4^{ij} = \frac{1}{2}\,\mathcal{I}_{ij}^{11}$ & ${U_{H}}_{ij}{U_{H}}_{ji}$ & $f_{8B}^{i} = \mathcal{I}_{i0}^{12}$ & ${U_{HL}}_{ii^\prime}{U_{L}}_{i^\prime j^\prime}{U_{LH}}_{j^\prime i}$ \\
\hline
\T\B $f_{4A}^{i} = \mathcal{I}_{i0}^{11}$ & ${U_{HL}}_{ii^\prime}{U_{LH}}_{i^\prime i}$ & \multicolumn{2}{c|}{  } \\
\hline
\hline
\multicolumn{2}{|c|}{ \T\B $\mathcal{O}(U^4)$ terms } & \multicolumn{2}{c|}{ \T\B $\mathcal{O}(U^6)$ terms }  \\
\hline
\T\B $f_{10}^{ijkl} = \frac{1}{4}\,\mathcal{I}_{ijkl}^{1111}$ & ${U_{H}}_{ij}{U_{H}}_{jk}{U_{H}}_{kl}{U_{H}}_{li}$ &  $f_{19}^{ijklmn} = \frac{1}{6} \,\mathcal{I}_{ijklmn}^{111111}$ & ${U_{H}}_{ij}{U_{H}}_{jk}{U_{H}}_{kl}{U_{H}}_{lm}{U_{H}}_{mn}{U_{H}}_{ni}$  \\
\hline
\T\B $f_{10A}^{ijk} = \mathcal{I}_{ijk0}^{1111}$ & ${U_{H}}_{ij}{U_{H}}_{jk}{U_{HL}}_{ki^\prime}{U_{LH}}_{i^\prime i}$ & $f_{19A}^{ijklm} = \mathcal{I}_{ijklm0}^{111111}$ & ${U_{H}}_{ij}{U_{H}}_{jk}{U_{H}}_{kl}{U_{H}}_{lm}{U_{HL}}_{mi^\prime}{U_{LH}}_{i^\prime i}$  \\
\hline
\T\B $f_{10B}^{ij} = \mathcal{I}_{ij0}^{112}$ & ${U_{H}}_{ij}{U_{HL}}_{ji^\prime}{U_{L}}_{i^\prime j^\prime}{U_{LH}}_{j^\prime i}$ & $f_{19B}^{ijkl} = \mathcal{I}_{ijkl0}^{11112}$ & ${U_{H}}_{ij}{U_{H}}_{jk}{U_{H}}_{kl}{U_{HL}}_{l i^\prime}{U_{L}}_{i^\prime j^\prime}{U_{LH}}_{j^\prime i}$  \\
\hline
\T\B $f_{10C}^{ij} = \frac{1}{2}\,\mathcal{I}_{ij0}^{112}$ & ${U_{HL}}_{ii^\prime}{U_{LH}}_{i^\prime j}{U_{HL}}_{j j^\prime}{U_{LH}}_{j^\prime i}$ & $f_{19C}^{ijkl} = \mathcal{I}_{ijkl0}^{11112}$ & ${U_{H}}_{ij}{U_{H}}_{jk}{U_{HL}}_{ki^\prime}{U_{LH}}_{i^\prime l}{U_{HL}}_{l j^\prime}{U_{LH}}_{j^\prime i}$  \\
\hline
\T\B $f_{10D}^{i} = \mathcal{I}_{i0}^{13}$ & ${U_{HL}}_{ii^\prime}{U_{L}}_{i^\prime j^\prime}{U_{L}}_{j^\prime k^\prime}{U_{LH}}_{k^\prime i}$ & $f_{19D}^{ijk} = \mathcal{I}_{ijk0}^{1113}$ & ${U_{H}}_{ij}{U_{H}}_{jk}{U_{HL}}_{k i^\prime}{U_{L}}_{i^\prime j^\prime}{U_{L}}_{j^\prime k^\prime}{U_{LH}}_{k^\prime i}$  \\
\hline
\multicolumn{2}{|c|}{ \T\B $\mathcal{O}(U^5)$ terms } & $f_{19E}^{ijkl} = \frac{1}{2}\,\mathcal{I}_{ijkl0}^{11112}$ & ${U_{H}}_{ij}{U_{HL}}_{j i^\prime}{U_{LH}}_{i^\prime k}{U_{H}}_{kl}{U_{HL}}_{l j^\prime}{U_{LH}}_{j^\prime i}$  \\
\hline
\T\B $f_{16}^{ijklm} = \frac{1}{5}\,\mathcal{I}_{ijklm}^{11111}$ & ${U_{H}}_{ij}{U_{H}}_{jk}{U_{H}}_{kl}{U_{H}}_{lm}{U_{H}}_{mi}$ & \multirow{2}{*}{$f_{19F}^{ijk} = \mathcal{I}_{ijk0}^{1113}$} & ${U_{H}}_{ij}{U_{HL}}_{j i^\prime}{U_{LH}}_{i^\prime k}{U_{HL}}_{k j^\prime}{U_{L}}_{j^\prime k^\prime}{U_{LH}}_{k^\prime i}$  \\
\cline{1-2}
\T\B $f_{16A}^{ijkl} = \mathcal{I}_{ijkl0}^{11111}$ & ${U_{H}}_{ij}{U_{H}}_{jk}{U_{H}}_{kl}{U_{HL}}_{l i^\prime}{U_{LH}}_{i^\prime i}$ & & $+{U_H}_{ij}{U_{HL}}_{ji'}{U_L}_{i'j'}{U_{LH}}_{j'k}{U_{HL}}_{kk'}{U_{LH}}_{k'i}$ \\
\hline
\T\B $f_{16B}^{ijk} = \mathcal{I}_{ijk0}^{1112}$ & ${U_{H}}_{ij}{U_{H}}_{jk}{U_{HL}}_{k i^\prime}{U_{L}}_{i^\prime j^\prime}{U_{LH}}_{j^\prime i}$ & $f_{19G}^{ij} = \mathcal{I}_{ij0}^{114}$ & ${U_{H}}_{ij}{U_{HL}}_{j i^\prime}{U_{L}}_{i^\prime j^\prime}{U_{L}}_{j^\prime k^\prime}{U_{L}}_{k^\prime l^\prime}{U_{LH}}_{l^\prime i}$  \\ 
\hline
\T\B $f_{16C}^{ijk} = \mathcal{I}_{ijk0}^{1112}$ & ${U_{H}}_{ij}{U_{HL}}_{j i^\prime}{U_{LH}}_{i^\prime k}{U_{HL}}_{k j^\prime}{U_{LH}}_{j^\prime i}$ & $f_{19H}^{ijk} = \frac{1}{3}\,\mathcal{I}_{ijk0}^{1113}$ & ${U_{HL}}_{i i^\prime}{U_{LH}}_{i^\prime j}{U_{HL}}_{j j^\prime}{U_{LH}}_{j^\prime k}{U_{HL}}_{k k^\prime}{U_{LH}}_{k^\prime i}$  \\
\hline
\T\B $f_{16D}^{ij} = \mathcal{I}_{ij0}^{113}$ & ${U_{H}}_{ij}{U_{HL}}_{j i^\prime}{U_{L}}_{i^\prime j^\prime}{U_{L}}_{j^\prime k^\prime}{U_{LH}}_{k^\prime i}$ &  $f_{19I}^{ij} = \mathcal{I}_{ij0}^{114}$ & ${U_{HL}}_{i i^\prime}{U_{LH}}_{i^\prime j}{U_{HL}}_{j j^\prime}{U_{L}}_{j^\prime k^\prime}{U_{L}}_{k^\prime l^\prime}{U_{LH}}_{l^\prime i}$ \\
\hline
\T\B $f_{16E}^{ij} = \mathcal{I}_{ij0}^{113}$ & ${U_{HL}}_{i i^\prime}{U_{LH}}_{i^\prime j}{U_{HL}}_{j j^\prime}{U_{L}}_{j^\prime k^\prime}{U_{LH}}_{k^\prime i}$ & $f_{19J}^{ij} = \frac{1}{2}\,\mathcal{I}_{ij0}^{114}$ & ${U_{HL}}_{i i^\prime}{U_{L}}_{i^\prime j^\prime}{U_{LH}}_{j^\prime j}{U_{HL}}_{j k^\prime}{U_{L}}_{k^\prime l^\prime}{U_{LH}}_{l^\prime i}$ \\
\hline
\T\B $f_{16F}^{i} = \mathcal{I}_{i0}^{14}$ & ${U_{HL}}_{i i^\prime}{U_{L}}_{i^\prime j^\prime}{U_{L}}_{j^\prime k^\prime}{U_{L}}_{k^\prime l^\prime}{U_{LH}}_{l^\prime i}$ & $f_{19K}^{i} = \mathcal{I}_{i0}^{15}$ & ${U_{HL}}_{i i^\prime}{U_{L}}_{i^\prime j^\prime}{U_{L}}_{j^\prime k^\prime}{U_{L}}_{k^\prime l^\prime}{U_{L}}_{l^\prime m^\prime}{U_{LH}}_{m^\prime i}$ \\
\hline
\end{tabular}
}
\end{center}
}
\caption{\it Universal coefficients for operators involving only $U$. There are 30 terms.  }
\label{tab:Uonly}
\end{table}

The expansion of the logarithm in Eq.~\ref{eq:extendedonelooplagrangian} may be written as
\begin{equation}
\mathcal{L}_\text{EFT}^{\text{1-loop}}[\phi] = -i c_s \text{tr} \sum_{n=1}^\infty \frac{1}{n} \int \frac{d^d q}{(2\pi)^d}  \left[ \left( q^2 - M^2 \right)^{-1} \left( 2q\cdot P - P^2 + U_H - U_{HL}\Delta_L^{-1} U_{LH} \right) \right]^n  \, ,
\label{eq:logexpansion}
\end{equation}
From this series, we can retrieve terms consisting of $P_\mu$ (dimension-1) and $U$ (dimension-1 or higher) up to certain dimensionality to obtain the universal expression of the UOLEA. In order to obtain all operators up to dimension six, one must evaluate terms in the series up to and including $n=6$, although this will generally also yield superfluous operators of dimension $d>6$.

As mentioned previously, an easier route to doing this uses covariant diagrams. Each universal operator structure in the expansion is represented by a covariant diagram, which can be written down directly by a systematic set of rules. We may then obtain the correct expression for the prefactor and master integrals associated with each operator structure without having to manually combine different related terms in the series of Eq.~\ref{eq:logexpansion}. Also, only the desired operator structures up to dimension six are retrieved, without superfluous higher-dimensional operators. Moreover, the diagrams involving adjacent contractions of covariant derivatives corresponding to $P^2$ terms can be excluded in this process as they are not necessary to determine the final UOLEA operator structures that are written in terms of $P$ commutators. 

As an example to illustrate this, we now calculate --- first without covariant diagrams --- the heavy-only universal coefficients associated to the expansion at order $n=4$, 
\begin{equation}
\left. \mathcal{L}_\text{EFT}^{\text{1-loop}}[\phi] \right|_{n=4} = - i c_s \frac{1}{4} \int \frac{d^d q}{(2\pi)^d} \text{tr} \left[ \left( q^2 - M^2 \right)^{-1} \left(2 q\cdot P - P^2 + U_H \right) \right]^4 \, .
\end{equation}
As discussed above, we shall consistently drop $P^2$ terms (though keeping them may serve as a cross-check). The resulting expansion, with the matrix indices written out explicitly, is given by 
\begin{align}
\left. \mathcal{L}_\text{EFT}^{\text{1-loop}}[\phi] \right|_{n=4} &= -i c_s \frac{1}{4} \int \frac{d^d q}{(4\pi)^d} \left( q^2 - M_i^2 \right)^{-1}\left( q^2 - M_j^2 \right)^{-1}\left( q^2 - M_k^2 \right)^{-1}\left( q^2 - M_l^2 \right)^{-1} \nonumber \\
& \mathrm{tr}\left[ {U_H}_{ij}{U_H}_{jk}{U_H}_{kl}{U_H}_{li} + 2 q_\mu P^\mu \delta_{ij} {U_H}_{jk} 2q_\rho P^\rho \delta_{kl} {U_H}_{li} \right. \nonumber \\
& \left. +{U_H}_{ij} 2q_\nu P^\nu \delta_{jk} {U_H}_{kl} 2q_\sigma P^\sigma \delta_{li} + 2^4 q_\mu q_\nu q_\rho q_\sigma P^\mu \delta_{ij} P^\nu \delta_{jk} P^\rho \delta_{kl} P^\sigma \delta_{li}  \right] \, .
\label{eq:foo1}
\end{align}
We can factor out the loop integrals by defining
\begin{align}
& \int\frac{d^dq}{(2\pi)^d} \frac{1}{(q^2-M_i^2)(q^2-M_j^2)(q^2-M_k^2)(q^2-M_l^2)} \equiv \,\mathcal{I}_{ijkl}^{1111} \\
& \int\frac{d^dq}{(2\pi)^d} \frac{q_{\mu}q_\nu}{(q^2-M_i^2)^2(q^2-M_j^2)^2}
\equiv g_{\mu\nu}\,\mathcal{I}[q^2]_{ij}^{22} \\
& \int\frac{d^dq}{(2\pi)^d} \frac{q_{\mu}q_\nu q_\rho q_\sigma}{(q^2-M_i^2)^4} \equiv \bigl(g_{\mu\nu}g_{\rho\sigma}+g_{\mu\rho}g_{\nu\sigma}+g_{\mu\sigma}g_{\nu\rho}\bigr)\,\mathcal{I}[q^4]_{i}^{4}
\end{align}
Eq.~\ref{eq:foo1} then becomes
\begin{align}
\left. \mathcal{L}_\text{EFT}^{\text{1-loop}}[\phi] \right|_{n=4} &= -i c_s \left[ \frac{1}{4}\,\mathcal{I}_{ijkl}^{1111} \,\mathrm{tr}\left( {U_H}_{ij}{U_H}_{jk}{U_H}_{kl}{U_H}_{li} \right) \right. \nonumber \\ 
& \left. +2\,\mathcal{I}[q^2]_{ij}^{22}\,\mathrm{tr}\left( P^\mu {U_H}_{ij} P_\mu {U_H}_{ji} \right) + 4 \,\mathcal{I}[q^4]_i^4\,\mathrm{tr} \left(P^\mu P^\nu P_\mu P_\nu \right) \right] \, .
\label{eq:n4example}
\end{align}
Finally, these may be put into the form of universal operator structures, where the covariant derivatives only enter in commutators, using 
\begin{equation}
\text{tr} \left( P^\mu U_H P_\mu U_H \right) \supset \frac{1}{2}\text{tr}\left[P^\mu, U_H\right]\left[P_\mu,U_H\right] \quad , \quad 
\text{tr}\left(P^\mu P^\nu P_\mu P_\nu \right) \supset \frac{1}{2} \left[ P^\mu, P^\nu\right]\left[P_\mu, P_\nu\right] \, , 
\label{eq:commutatorrelation}
\end{equation}
again neglecting $P^2$ terms from adjacent contracted covariant derivatives~\footnote{Keeping the $P^2$ terms in the derivation and the identities (\ref{eq:commutatorrelation}) just ensures the exact equality of the relations between the commutator and non-commutator operator structures, but we see that the non-$P^2$ terms are sufficient for determining this.}. Comparing Eqs.~\ref{eq:n4example} and~\ref{eq:commutatorrelation}, we obtain
\begin{align}
\mathcal{L}_\text{EFT}^{\text{1-loop}}[\phi] \supset 
& -i c_s \bigl[ f_{10}^{ijkl}\,\mathrm{tr}\left( {U_H}_{ij}{U_H}_{jk}{U_H}_{kl}{U_H}_{li} \right) \nonumber \\ 
& +f_7^{ij} \,\mathrm{tr}\left( [P^\mu, {U_H}_{ij}] [P_\mu, {U_H}_{ji}] \right) 
+f_3^{i}\,\mathrm{tr} \left( {G^\prime}^{\mu\nu}_i{G^\prime_{\mu\nu}}_i\right) \bigr]
\, ,
\end{align}
with
\begin{equation}
f_{10}^{ijkl} \equiv \frac{1}{4}\,\mathcal{I}_{ijkl}^{1111} \quad , \quad
f_7^{ij} \equiv \mathcal{I}[q^2]_{ij}^{22} \quad , \quad 
f_3^i \equiv 2 \,\mathcal{I}[q^4]_i^4 \quad ,
\end{equation}
and we defined $G^\prime_{\mu\nu}\equiv-[P_\mu,P_\nu]=-igG_{\mu\nu}$. In this way we have obtained the universal coefficients $f_{10}^{ijkl}, f_7^{ij}, f_3^i$ in the heavy-only UOLEA.

While this simple example was chosen for clarity, in general the combinatorics and tedious manipulations involved can grow rapidly as we compute operators with more $P$'s and $U$'s, especially when additional structures such as the heavy-light ones are added. We need a book-keeping device to streamline the calculation, and such a device is provided by covariant diagrams.

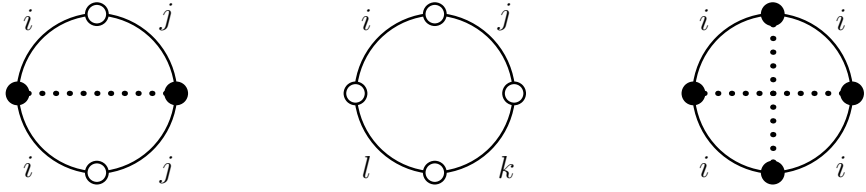
\begin{figure}[tbp]
\begin{center}
\begin{fmffile}{n4cd}
\begin{fmfgraph*}(60,60)
\fmfsurround{vP2,vU1,vP1,vU2}
\fmf{plain,left=0.4,label=$i$,l.d=4pt}{vU2,vP1,vU1}
\fmf{plain,left=0.4,label=$j$,l.d=4pt}{vU1,vP2,vU2}
\fmfv{decor.shape=circle,decor.filled=full,decor.size=4thick}{vP1,vP2}
\fmfv{decor.shape=circle,decor.filled=empty,decor.size=4thick}{vU1,vU2}
\fmf{dots,width=thick}{vP1,vP2}
\end{fmfgraph*}
\hspace{60pt}
\begin{fmfgraph*}(60,60)
\fmfsurround{vU3,vU2,vU1,vU4}
\fmf{plain,left=0.4,label=$i$,l.d=4pt}{vU1,vU2}
\fmf{plain,left=0.4,label=$j$,l.d=4pt}{vU2,vU3}
\fmf{plain,left=0.4,label=$k$,l.d=4pt}{vU3,vU4}
\fmf{plain,left=0.4,label=$l$,l.d=4pt}{vU4,vU1}
\fmfv{decor.shape=circle,decor.filled=empty,decor.size=4thick}{vU1,vU2,vU3,vU4}
\end{fmfgraph*}
\hspace{60pt}
\begin{fmfgraph*}(60,60)
\fmfsurround{vP3,vP2,vP1,vP4}
\fmf{plain,left=0.4,label=$i$,l.d=4pt}{vP1,vP2,vP3,vP4,vP1}
\fmfv{decor.shape=circle,decor.filled=full,decor.size=4thick}{vP1,vP2,vP3,vP4}
\fmf{dots,width=thick}{vP1,vP3}
\fmf{dots,width=thick}{vP2,vP4}
\end{fmfgraph*}
\end{fmffile}
\end{center}
\caption{Covariant diagrams for the covariant derivative expansion at order $n=4$. }
\label{fig:n4covariantdiagrams}
\end{figure}

Using covariant diagrams we could have just written down Eq.~\ref{eq:n4example} directly, without explicitly going through the steps detailed above.
According to the covariant diagram rules set out in \cite{covariantdiagrams}, we simply draw all possible loops joining up 4 filled or empty circles by solid lines, representing $P$ and $U_H$ respectively, and ensure that each filled circle is also joined up by a dotted line to another filled circle, representing Lorentz contractions between two $P$'s. There are 3 topologically-independent possibilities of drawing such diagrams (not counting those with adjacent $P_\mu$ contractions since we always drop $P^2$ terms), which we show in Fig.~\ref{fig:n4covariantdiagrams}. The associated mathematical expression corresponding to Eq.~\ref{eq:n4example} can then be read off directly according to the rules.~\footnote{When first introduced to covariant diagrams, it can be tempting to skeptically assume that the representation of the expansion in terms of diagrams is trivial, a pictorial representation or visual shorthand for essentially the same algebraic computation. In fact, direct experience with calculations using previous methods of evaluating the expansion confirms that this diagrammatic approach is not merely a replacement of the same steps by dots and lines, but represents a genuine simplification that renders previous steps unnecessary. This is analogous to how Feynman diagrams for correlation functions simplify the process of setting up perturbative expansions and their Wick contractions. }  

This method has been applied to rederive the heavy-only UOLEA in a simpler form in Ref.~\cite{covariantdiagrams}, shown in Eq.~\ref{eq:heavyonlyUOLEA}; we also give these results in the following tables for completeness. We have calculated the covariant diagrams corresponding to all possible vertex insertions with additional vertices due to the heavy-light terms in Eq.~\ref{eq:extendedonelooplagrangian}. As a cross-check, the calculation has also been incorporated into a code that we are developing to automatically generate the diagrams and solve for the corresponding universal coefficients of the commutated operator structure of the UOLEA. 

An additional cross-check can be found in the resulting structure of the coefficients, where we noticed that the heavy-light coefficients are related to the heavy-only universal ones by making appropriate replacements of indices corresponding to heavy fields with indices for light fields. Given the arbitrary choice of making a field heavy or light, this interrelated structure was to be expected, but has previously been obscured by the obliqueness of past methods for treating heavy-light matching. Here we see this appear explicitly from our computation.

For example, starting from
\begin{equation}
f_7^{ij}\,\mathrm{tr}\left( [P^\mu, {U_H}_{ij}] [P_\mu, {U_H}_{ji}] \right) = \mathcal{I}[q^2]_{ij}^{22}\,\mathrm{tr}\left( [P^\mu, {U_H}_{ij}] [P_\mu, {U_H}_{ji}] \right)
\end{equation}
we can substitute either $i$ or $j$ by a light field $i'$ to arrive at
\begin{equation}
f_{7A}^{i}\,\mathrm{tr}\left( [P^\mu, {U_{HL}}_{ii'}] [P_\mu, {U_{LH}}_{i'i}] \right) = 2\,\mathcal{I}[q^2]_{i0}^{22}\,\mathrm{tr}\left( [P^\mu, {U_{HL}}_{ii'}] [P_\mu, {U_{LH}}_{i'i}] \right)\,, 
\end{equation}
with $\,\mathcal{I}[q^2]_{i0}^{22}$ defined by
\begin{equation}
\int\frac{d^dq}{(2\pi)^d} \frac{q_{\mu}q_\nu}{(q^2-M_i^2)^2(q^2)^2}
\equiv g_{\mu\nu}\,\mathcal{I}[q^2]_{i0}^{22}\,.
\end{equation}
This algorithmic procedure of obtaining heavy-light universal coefficients from heavy-only ones holds in general, as can be seen from the covariant diagram rules. This observation potentially allows the future UOLEA extensions discussed in Section~\ref{sec:completeUOLEA} to be streamlined. 

The universal coefficients are listed in Tables~\ref{tab:PonlyandU2P2andU1P4}, \ref{tab:U3P2}, \ref{tab:U2P4}, \ref{tab:U4P2}, \ref{tab:Uonly}. They are grouped together and tabulated according to their universal operator structures. Coefficients and operators involving $\mathcal{O}(U^2 P^2)$, $\mathcal{O}(U P^4)$ and those containing only $P$'s can be found in Table~\ref{tab:PonlyandU2P2andU1P4}. The $\mathcal{O}(U^3 P^2)$ terms are shown in Table~\ref{tab:U3P2}, while the $\mathcal{O}(U^2 P^4)$ terms are in Table~\ref{tab:U2P4}, with $\mathcal{O}(U^4 P^2)$ terms in Table~\ref{tab:U4P2}. Finally the $U$-only terms are in Table~\ref{tab:Uonly}. Each of the coefficients appearing in these tables $f_N$ is expressed in terms of combinations of master integrals $\mathcal{I}$, which are defined as 
\begin{equation}
\int\frac{d^dq}{(2\pi)^d} \frac{q^{\mu_1}\cdots q^{\mu_{2n_c}}}{(q^2-M_i^2)^{n_i}(q^2-M_j^2)^{n_j}\cdots (q^2)^{n_L}}
\,\equiv\, g^{\mu_1\dots\mu_{2n_c}} \,\mathcal{I}[q^{2n_c}]_{ij\dots 0}^{n_i n_j\dots n_L}
\end{equation}
where $g^{\mu_1\dots\mu_{2n_c}}$ is the completely symmetric tensor, e.g.\ $g^{\mu\nu\rho\sigma}=g^{\mu\nu}g^{\rho\sigma} +g^{\mu\rho}g^{\nu\sigma} +g^{\mu\sigma}g^{\nu\rho}$. Note that the $U$ matrix indices for the parts involving light fields are denoted by primed indices, which are contracted together independently of the unprimed heavy indices (that also appear in $f_N$). 
The master integrals are discussed in greater detail in Appendix~\ref{Integrals.APP}. 

Complete expressions of the universal coefficients $f_N$ in terms of heavy particle masses are available in a \texttt{Mathematica} notebook in the arXiv submission. In applications of integrating out a single heavy field, or a set of degenerate heavy fields, these expressions become particularly simple, which we provide in Appendix~\ref{app:uc}.

\section{Example: Integrating out a singlet scalar}
\label{sec:example}

We now apply the extended UOLEA to integrate out a heavy real singlet scalar $\phi$ coupling to the light SM Higgs doublet $H$. The Lagrangian contains 
\begin{equation}
\mathcal{L} \supset \frac{1}{2}\left(\partial_\mu\phi\right)^2 - \frac{1}{2}m_\phi^2\phi^2 - A|H|^2\phi -\frac{1}{2}\kappa |H|^2 \phi^2 - \frac{1}{3!} \mu \phi^3 - \frac{1}{4!} \lambda_\phi \phi^4 -\frac{1}{2}\lambda_H|H|^4\, .
\end{equation}
Since no $\mathbb{Z}_2$ symmetry is assumed, $\phi$ can couple linearly to the Higgs and so can give rise to tree-level contributions to matching. This is obtained for example in Ref.~\cite{HLM}, where they also give the heavy-only one-loop matching in a choice of UV parameter renormalisation such that the scheme-dependent finite terms are set to zero~\footnote{In practice, this is equivalent to just setting the solution to the classical equation of motion for the heavy field to zero.}. Here we perform the one-loop matching in the $\overline{MS}$ renormalisation scheme and include the heavy-light contributions, which to our knowledge has not appeared previously in the literature.  

The equation of motion for $\phi$ up to $\mathcal{O}(1/m_\phi^8)$ can be solved order by order to give 
\begin{align}
\phi_c &= -\frac{A}{m_\phi^2}|H|^2 + \left(\frac{\kappa A}{m_\phi^4} - \frac{1}{2}\frac{\mu A^2}{m_\phi^6}\right)|H|^4 + \frac{A}{m_\phi^4}\partial^2|H|^2 \nonumber \\
& \quad\quad - \left( \frac{\kappa^2 A}{m_\phi^6} - \frac{1}{2}\frac{\kappa\mu A^2}{m_\phi^8} - \frac{1}{6}\frac{\lambda_\phi A^3}{m_\phi^8} \right) |H|^6 - \frac{\kappa A}{m_\phi^6} |H|^2 \partial^2 |H|^2 + \text{...} \, ,
\end{align}
where we have dropped $\partial^4 |H|^2$ and $\partial^2 |H|^4$ terms that can only contribute as a total derivative since they are already of dimension 6. 

Next we write the multiplet for the heavy field $\phi$ and light field $H$ as $\Phi = (\phi, H, \tilde{H})$, where we separate the complex Higgs doublet into $H$ and $\tilde{H} \equiv i\sigma_2 H^*$ so as to obtain the same $c_s = 1/2$ factor in the path integral for $\Phi$. The quadratic term in the Lagrangian is then of the form 
\begin{equation}
\mathcal{L} \supset \frac{1}{2}\, \Phi^\dagger \twomatrix{ P^2 - m_\phi^2 - U_{\phi} & -\left({\bf U}_{\phi H}\right)_{1\times2}  \\ -\left({\bf U}_{H\phi}\right)_{2\times1} & \left(P^2 - m_H^2 -{\bf U}_{HH}\right)_{2\times2} } \Phi \, ,
\end{equation}
where
{\small
\begin{align}
&U_\phi = \kappa |H|^2 + \mu \phi_c + \frac{1}{2}\lambda_\phi \phi_c^2 \, , \nonumber \\
&\left({\bf U}_{\phi H}\right)_{1\times 2} = \left( AH^\dagger + \kappa \phi_c H^\dagger \, , \, A \tilde{H}^\dagger + \kappa \phi_c \tilde{H}^\dagger \right) \, , \nonumber \\
&\left({\bf U}_{H\phi}\right)_{2\times 1} = \column{ AH + \kappa H \phi_c \\ A \tilde{H} + \kappa \tilde{H} \phi_c } \, , \nonumber \\
\left({\bf U}_{HH}\right)_{2\times2} = &\twomatrix{ (A\phi_c + \frac{1}{2}\kappa\phi_c^2)\identity_2 + \lambda_H \left(|H|^2\identity_2 + HH^\dagger\right) & \lambda_H H\tilde{H}^\dagger \\ \lambda_H \tilde{H}H^\dagger & (A\phi_c + \frac{1}{2}\kappa\phi_c^2)\identity_2 + \lambda_H \big(|H|^2\identity_2 + \tilde{H}\tilde{H}^\dagger\big)} \, . 
\label{eq:Umatrix}
\end{align}
}

The heavy-only part of the one-loop matching contributes to the Wilson coefficients of the operators
\begin{equation}
\mathcal{O}_6 = |H|^6 \quad , \quad \mathcal{O}_H = \frac{1}{2}\left(\partial_\mu |H|^2 \right)^2 \, .
\end{equation}
We will focus on obtaining these two operators. It is straightforward to identify the UOLEA terms that can contribute to an operator structure involving six $H$'s, and another with two $P$'s and four $H$'s, by counting operator dimensions. Since we are including heavy-light loops in the matching the following operators at $\mathcal{O}(P^2 H^4)$ may also in principle be generated,
\begin{equation}
\mathcal{O}_R = |H|^2 |D_\mu H|^2 \quad , \quad \mathcal{O}_T = \frac{1}{2}\left(H^\dagger \overleftrightarrow{D}_\mu H \right)^2 \, .
\end{equation}
However we expect to get a vanishing coefficient for $\mathcal{O}_T$ as $\phi$ is a singlet with zero hypercharge and so cannot break custodial symmetry. 

Considering the smallest operator dimensions in the heavy, light, and heavy-light entries of the $U$ matrix in Eq.~\ref{eq:Umatrix}, we may isolate the following coefficients to compute for $\mathcal{O}_6$:
\begin{equation}
f_2, f_4, f_{4A}, f_8, f_{8A}, f_{8B}, f_{10A}, f_{10B}, f_{10C}, f_{10D}, f_{16C}, f_{16E}, f_{19H} \, .
\end{equation}
Taking the trace of the corresponding universal operator structures given in Table~\ref{tab:Uonly} with the $U$ matrix of Eq.~\ref{eq:Umatrix} we get the following Wilson coefficient of $\mathcal{O}_6$ expressed in terms of the rescaled universal coefficients $\tilde{f}_N\equiv f_N/\frac{i}{16\pi^2}$, \\
\resizebox{\linewidth}{!}{
\begin{minipage}{\linewidth}
\begin{align}
\nonumber \mathcal{L}_\text{EFT}^{\text{1-loop}}[\phi] &\supset \frac{1}{2(4\pi)^2} \Bigg\{ 
\tilde{f}_2 \Bigg( -\frac{\kappa A(A \lambda_\phi + \kappa \mu)}{m_\phi^6} + \frac{\mu A^2(4A \lambda_\phi + 3 \kappa \mu)}{6m_\phi^8} \Bigg) \\
\nonumber &+ \tilde{f}_4 \Bigg( \frac{\kappa A(A \lambda_\phi + 2\kappa \mu)}{m_\phi^4} - \frac{\mu A^2(A \lambda_\phi + 3\kappa \mu)}{m_\phi^6} + \frac{\mu^3 A^3}{m_\phi^8}\Bigg) 
+ \tilde{f}_{4A} \Bigg( \frac{6\kappa^2 A^2}{m_\phi^4} - \frac{2\kappa \mu A^3}{m_\phi^6} + \frac{\mu^3 A^3}{m_\phi^8}\Bigg) \\
\nonumber&+\tilde{f}_8\Bigg( \kappa^3 -\frac{3\kappa^2 \mu A}{m_\phi^2} + \frac{3A^2 \kappa \mu^2}{m_\phi^4} - \frac{A^3\mu^3}{m_\phi^6}\Bigg) 
+ \tilde{f}_{8A}\Bigg( -\frac{4A^2\kappa^2}{m_\phi^2} + \frac{A^3(A \lambda_\phi + 6 \kappa \mu)}{m_\phi^4} - \frac{\mu^2 A^4}{m_\phi^6}\Bigg) \\
\nonumber&+\tilde{f}_{8B}\Bigg( -\frac{12\lambda_H A^2 \kappa}{m_\phi^2} + \frac{7A^4\kappa}{m_\phi^4} - \frac{\mu A^5}{m_\phi^6}\Bigg) 
+\tilde{f}_{10A}\Bigg( 2A^2\kappa^2 - \frac{4\kappa \mu A^3}{m_\phi^2} + \frac{2A^4\mu^2}{m_\phi^4}\Bigg) \\
\nonumber &+ \tilde{f}_{10B} \Bigg( 6\lambda_H A^2 \kappa - \frac{2A^3(3\lambda_H \mu + \kappa A)}{m_\phi^2} + \frac{2\mu A^5}{m_\phi^4}\Bigg) 
+ \tilde{f}_{10C} \Bigg( -\frac{16A^4 \kappa}{m_\phi^2}\Bigg) \\
\nonumber &+ \tilde{f}_{10D}\Bigg( 18\lambda_H^2 A^2 - \frac{12A^4 \lambda_H}{m_\phi^2} +\frac{2A^6}{m_\phi^4}\Bigg) 
+ \tilde{f}_{16C} \Bigg( 4\kappa A^4 - \frac{4A^5 \mu}{m_\phi^2}\Bigg) 
+ \tilde{f}_{16E}\Bigg( 12\lambda_H A^4 - \frac{4A^6}{m_\phi^2}\Bigg) \\
&+ \tilde{f}_{19H} \left( 8A^6\right) \Bigg\} \,{\cal O}_6 \,.
\end{align}
\end{minipage}
}
Here for simplicity, indices on the universal coefficients $\tilde{f}_N$ have been omitted, since all of them take $\phi$, the only heavy field in the theory. Explicit expressions for these degenerate universal coefficients in terms of the heavy particle mass can be found in Appendix~\ref{app:uc}.

We may similarly isolate the following coefficients to compute for $\mathcal{O}_H$:
\begin{equation}
f_2, f_4, f_{4A}, f_7, f_{7A}, f_{8A}, f_{8B}, f_{11B}, f_{11C}, f_{11D}, f_{17J}, f_{18E} \, .
\end{equation}
Here the universal operator structures are given in Tables~\ref{tab:U3P2}, \ref{tab:U4P2}, \ref{tab:Uonly}. Taking the trace of these matrices we find the following combinations of universal coefficients for the Wilson coefficient of $\mathcal{O}_H$, \\
\resizebox{\linewidth}{!}{
\begin{minipage}{\linewidth}
\begin{align}
\mathcal{L}_\text{EFT}^{\text{1-loop}}[\phi] \supset \frac{1}{2(4\pi)^2}  & \Bigg\{ 
\tilde{f}_{2}\Bigg( \frac{2 A  \left(A \lambda _{\phi } +\kappa  \mu \right)}{m_{\phi}^6} \Bigg)
+ \tilde{f}_{4}\Bigg(\frac{4 A \mu  \left(A \mu -\kappa  m_{\phi}^2\right)}{m_{\phi}^6} \Bigg)
- \tilde{f}_{4A}\Bigg(\frac{8 A^2 \kappa }{m_{\phi}^4} \Bigg)
\nonumber \\
&
-\tilde{f}_{7}\left(\frac{2  \left(\kappa  m_{\phi}^2-A \mu \right)^2}{m_{\phi}^4}\right)
+\tilde{f}_{7A}\left(\frac{4 A^2 \kappa }{m_{\phi}^2}\right)
-\tilde{f}_{8A}\left(\frac{4 A^3 \mu }{m_{\phi}^4}\right)
-\tilde{f}_{8B}\left(\frac{4 A^4 }{m_{\phi}^4}\right)
\nonumber \\
&
+ \tilde{f}_{11B} \left(\frac{4A^3 \mu }{m_{\phi}^2} -4A^2\kappa \right)
-\tilde{f}_{11C} \left(2A^2  \lambda_H \right)
- \tilde{f}_{11D} \left( 8 A^2 \lambda_H -\frac{4A^4}{ m_{\phi}^2} \right)
\nonumber \\
&
- \tilde{f}_{17J} \left(2 A^4\right)
- \tilde{f}_{18E} \left(4 A^4\right)
\Bigg\} \,{\cal O}_H \,.
\label{eq:OH}
\end{align}
\end{minipage}
}

For $\mathcal{O}_R$ only the following structures are necessary,
\begin{equation}
f_{7A}, f_{11A}, f_{11C}, f_{11D}, f_{17I} \, .
\end{equation}
These are listed in Tables~\ref{tab:U3P2}, \ref{tab:U4P2}, \ref{tab:Uonly} and yield the following result,
\begin{align}
\mathcal{L}_\text{EFT}^{\text{1-loop}}[\phi] \supset  \frac{1}{2(4\pi)^2 } & \Bigg\{ 
\tilde{f}_{7A}\left(\frac{4 A^2 \kappa }{m_{\phi}^2}\right)
+ \tilde{f}_{11A} \left(\frac{2A^3 \mu }{m_{\phi}^2}-2A^2\kappa \right)
+\tilde{f}_{11C} \left(\frac{2A^4}{m_{\phi}^2}- 2A^2\lambda _h\right)
\nonumber \\
&
- \tilde{f}_{11D} \left(8 A^2 \lambda _h \right) 
- \tilde{f}_{17I}  \left(4 A^4\right)
\Bigg\} \,{\cal O}_R \, .
\label{eq:OR}
\end{align}

As expected, there are no contributions to $\mathcal{O}_T$.

This simple example serves to outline the steps involved in using the UOLEA to calculate heavy-light matching at one-loop. The phenomenology of real singlet scalar extensions of the Standard Model has been studied in e.g.~\cite{singletpheno} and their one-loop structure could be relevant for future precision measurements~\cite{future}.

\section{Conclusion}
\label{sec:conclusion}

To the casual reader, it might appear that Feynman diagram matching is still quicker than using functional methods (at least for now). But this is only due to familiarity with a well-developed set of standard results and automated tools. Indeed, when using Feynman diagrams, it is unnecessary to always start the calculation from the beginning, setting up correlation functions then performing Wick contractions by hand. There is a similar redundancy in evaluating the covariant derivative expansion of the one-loop path integral, a fact obscured by the many methods for doing this in unnecessarily complicated ways. The lack of a simple textbook standard and the perceived limitations of its applicability in including heavy-light loops may have contributed to the limited adoption of functional methods in one-loop matching for practical purposes. 

As summarised in Section~\ref{sec:introduction}, recent developments have led to such a standardised and systematic method of accounting for the terms in the one-loop path integral -- the covariant diagrams~\cite{covariantdiagrams} -- that serve the same purpose for the expansion of the one-loop path integral as Feynman diagrams do for the perturbative expansion of correlation function amplitudes. It had also previously been noticed that this one-loop path-integral expansion can be evaluated model-independently up to any finite order to give a {\it universal expression} -- the UOLEA~\cite{HLM, UOLEA}. The obvious step is then to do this once-and-for-all.

In this work we have further developed the UOLEA by including the universal terms necessary for matching with heavy-light loops. Our key results are the universal coefficients presented in Tables~\ref{tab:PonlyandU2P2andU1P4}, \ref{tab:U3P2}, \ref{tab:U2P4}, \ref{tab:U4P2}, \ref{tab:Uonly}. Their explicit expressions in terms of heavy particle masses can be found in a \texttt{Mathematica} notebook in the arXiv submission, whose degenerate limits are collected in Appendix~\ref{app:uc}. We have demonstrated how to use these universal results to efficiently compute EFT operator coefficients with a singlet scalar model example. In future work~\cite{workinprogress} we plan to complete the UOLEA by including all possible structures one may encounter in evaluating the covariant derivative expansion, to provide a standard set of results that can serve as a reference for one-loop matching.

\subsubsection*{Acknowledgements}
SE and ZZ would like to thank the DESY Theory Group for hospitality. 
TY is supported by a Junior Research Fellowship from Gonville and Caius College, Cambridge. 
The work of SE and ZZ is supported in part by the U.S. Department of Energy under grant DE-SC0007859. 
The work of JQ was supported by the UK STFC Grant ST/L000326/1.

\appendix

\section{Master integrals}
\label{Integrals.APP}

The universal coefficients $f_N$ presented in this paper are written in terms of master integrals $\,\mathcal{I}$, defined by
\begin{equation}
\int\frac{d^dq}{(2\pi)^d} \frac{q^{\mu_1}\cdots q^{\mu_{2n_c}}}{(q^2-M_i^2)^{n_i}(q^2-M_j^2)^{n_j}\cdots (q^2)^{n_L}}
\,\equiv\, g^{\mu_1\dots\mu_{2n_c}} \,\mathcal{I}[q^{2n_c}]_{ij\dots 0}^{n_i n_j\dots n_L} \,.
\end{equation}
With the following reduction formulas, 
\begin{align}
\mathcal{I}[q^{2n_c}]_{ij\dots0}^{n_in_j\dots n_L} &= \frac{1}{\Delta_{ij}^2} \bigl(\,\mathcal{I}[q^{2n_c}]_{ij\dots0}^{n_i,n_j-1,\dots n_L} -\,\mathcal{I}[q^{2n_c}]_{ij\dots0}^{n_i-1,n_j\dots n_L}\bigr) \,,\\
\mathcal{I}[q^{2n_c}]_{ij\dots0}^{n_in_j\dots n_L} &= \frac{1}{M_i^2} \bigl(\,\mathcal{I}[q^{2n_c}]_{ij\dots0}^{n_in_j\dots, n_L-1} -\,\mathcal{I}[q^{2n_c}]_{ij\dots0}^{n_i-1,n_j\dots n_L}\bigr) \,,\\
\mathcal{I}[q^{2n_c}]_{ij\dots0}^{n_in_j\dots n_L} &= \frac{1}{n_i-1} \frac{\partial}{\partial M_i^2} \,\mathcal{I}[q^{2n_c}]_{ij\dots0}^{n_i-1,n_j\dots n_L} \,,
\end{align}
where $\Delta_{ij}^2\equiv M_i^2-M_j^2$, it can be shown that
\begin{align}
\mathcal{I}[q^{2n_c}]_{ij\dots0}^{n_in_j\dots n_L} =& \sum_{p_i=0}^{n_i-1} \left[\, \frac{1}{p_i!} \left(\frac{\partial}{\partial M_i^2}\right)^{p_i} \frac{1}{(M_i^2)^{n_L} (\Delta_{ij}^2)^{n_j} (\Delta_{ik}^2)^{n_k} \dots} \,\right] \,\mathcal{I}[q^{2n_c}]_i^{n_i-p_i} \nonumber\\
&+ \sum_{p_j=0}^{n_j-1} \left[\, \frac{1}{p_j!} \left(\frac{\partial}{\partial M_j^2}\right)^{p_j} \frac{1}{(M_j^2)^{n_L} (\Delta_{ji}^2)^{n_j} (\Delta_{jk}^2)^{n_k} \dots} \,\right] \,\mathcal{I}[q^{2n_c}]_j^{n_j-p_j}+ \dots 
\label{eq:MIred}
\end{align}
With Eq.~\ref{eq:MIred}, any master integral $\,\mathcal{I}$ can be decomposed into a sum of heavy-only degenerate master integrals of the form $\,\mathcal{I}[q^{2n_c}]_i^{n_i}$, for which the general expression reads
\begin{equation}
\mathcal{I}[q^{2n_c}]_i^{n_i} = \frac{i}{16\pi^2} \bigl(-M_i^2\bigr)^{2+n_c-n_i}
\frac{1}{2^{n_c}(n_i-1)!} \frac{\Gamma(\frac{\epsilon}{2}-2-n_c +n_i)}{\Gamma(\frac{\epsilon}{2})} \Bigl(\frac{2}{\epsilon} -\gamma +\log 4\pi-\log\frac{M_i^2}{\mu^2}\Bigr) \,,
\end{equation}
where $d=4-\epsilon$ is the spacetime dimension, and $\mu$ is the renormalization scale. A table of $\,\mathcal{I}[q^{2n_c}]_i^{n_i}$ for various $n_c$ and $n_i$ can be found in Appendix~A of~\cite{covariantdiagrams}.

\section{Explicit expressions for universal coefficients with degenerate heavy fields}
\label{app:uc}

In the specific case where only one heavy field is being integrated out, such as in the example above with a real singlet scalar, or when all the heavy fields are degenerate, the universal coefficients listed in Tables~\ref{tab:PonlyandU2P2andU1P4}, \ref{tab:U3P2}, \ref{tab:U2P4}, \ref{tab:U4P2}, \ref{tab:Uonly} take a simple form. We list below each of the coefficients with the master integrals written out explicitly for degenerate heavy fields~\footnote{The heavy-only degenerate UOLEA coefficients were first obtained in Ref.~\cite{HLM}.}. The notation employed is the following:
\begin{equation}
\nonumber \tilde{f}_N = -i \,16 \pi^2 \,f_N,  ~~~~~~~ \tilde{\mathcal{I}}[q^{2n_c}]_{i0}^{n_i n_L} = -i \,16 \pi^2 \,\mathcal{I}[q^{2n_c}]_{i0}^{n_i n_L} \ ,
\end{equation}
and the coefficients are:
\vspace{-24pt}
\begin{multicols}{2}
\raggedcolumns
\begin{scriptsize}
\begin{align*}
\tilde{f}_2^{i} &= \,\tilde{\mathcal{I}}_{i}^{1}= M_i^2\left( 1-\log \frac{M_i^2}{\mu^2}\right) \\
\tilde{f}_3^{i} &= 2\,\tilde{\mathcal{I}}[q^4]_{i}^{4} = -\frac{1}{12}\log \frac{M_i^2}{\mu^2}  \\
\tilde{f}_4^{i} &= \frac{1}{2}\,\tilde{\mathcal{I}}_{i}^{2} = -\frac{1}{2}\log\frac{M_i^2}{\mu^2} \\
\tilde{f}_{4A}^{i} &= \,\tilde{\mathcal{I}}_{i0}^{11}= 1-\log \frac{M_i^2}{\mu^2} \\
\tilde{f}_5^{i} &= 16\,\tilde{\mathcal{I}}[q^6]_{i}^{6} = -\frac{1}{60 M_i^2} \\
\tilde{f}_6^{i} &= (32/3)\,\tilde{\mathcal{I}}[q^6]_{i}^{6} = -\frac{1}{90 M_i^2}  \\
\tilde{f}_7^{i} &= \tilde{\mathcal{I}}[q^2]_{i}^{4}= - \frac{1}{12M_i^2} \\
\tilde{f}_{7A}^{i} &= 2\,\tilde{\mathcal{I}}[q^2]_{i0}^{22} = -\frac{1}{2M_i^2} \\
\tilde{f}_{8}^{i} &= \frac{1}{3}\,\tilde{\mathcal{I}}_{i}^{3} = -\frac{1}{6 M_i^2} \\ 
\tilde{f}_{8A}^{i} &= \,\tilde{\mathcal{I}}_{i0}^{21} = -\frac{1}{M_i^2} \\
\tilde{f}_{8B}^{i} &= \,\tilde{\mathcal{I}}_{i0}^{12} = \frac{1}{M_i^2}\left( 1 -\log \frac{M_i^2}{\mu^2}\right) \\
\tilde{f}_9^{i} &= 8\,\tilde{\mathcal{I}}[q^4]_{i}^{5} = -\frac{1}{12 M_i^2}  \\
\tilde{f}_{10}^{i} &= \frac{1}{4}\,\tilde{\mathcal{I}}_{i}^{4} = \frac{1}{24 M_i^4} \\
\\
\\
\tilde{f}_{10A}^{i} &= \,\tilde{\mathcal{I}}_{i0}^{31} = \frac{1}{2 M_i^4} \\
\tilde{f}_{10B}^{i} &= \,\tilde{\mathcal{I}}_{i0}^{22} = -\frac{1}{M_i^4}\left( 2 - \log \frac{M_i^2}{\mu^2}\right) \\
\tilde{f}_{10C}^{i} &= \frac{1}{2}\,\tilde{\mathcal{I}}_{i0}^{22} = -\frac{1}{2 M_i^4}\left( 2 - \log \frac{M_i^2}{\mu^2}\right) \\
\tilde{f}_{10D}^{i} &= \,\tilde{\mathcal{I}}_{i0}^{13} = \frac{1}{M_i^4}\left( 1 - \log \frac{M_i^2}{\mu^2}\right) \\
\tilde{f}_{11}^{i} &= 4\,\tilde{\mathcal{I}}[q^2]^{5}_{i}  = \frac{1}{12 M_i^4} \\
\tilde{f}_{11A}^{i} &= 4\,\tilde{\mathcal{I}}[q^2]_{i0}^{32} = \frac{1}{2 M_i^4} \\
\tilde{f}_{11B}^{i} &= 2\left(\,\tilde{\mathcal{I}}[q^2]_{i0}^{41} + \,\tilde{\mathcal{I}}[q^2]_{i0}^{32}\right) =  \frac{1}{3 M_i^4} \\
\tilde{f}_{11C}^{i} &= 4\,\tilde{\mathcal{I}}[q^2]_{i0}^{23} = -\frac{1}{M_i^4}\left( \frac{5}{2} - \log \frac{M_i^2}{\mu^2}\right)  \\
\tilde{f}_{11D}^{i} &= 2\left(\,\tilde{\mathcal{I}}[q^2]_{i0}^{14} + \,\tilde{\mathcal{I}}[q^2]_{i0}^{23}\right) = -\frac{1}{2M_i^4}  \\
\tilde{f}_{12}^{i} &= 4\,\tilde{\mathcal{I}}[q^4]_{i}^{6} = \frac{1}{120 M_i^4} \\
\tilde{f}_{12A}^{i} &= 8\,\tilde{\mathcal{I}}[q^4]_{i0}^{33} = \frac{1}{6 M_i^4} \\
\tilde{f}_{13}^{i} &= 20\,\tilde{\mathcal{I}}[q^4]_{i}^{6} = \frac{1}{24 M_i^4} \\
\tilde{f}_{13A}^{i} &= 4\left(\,\tilde{\mathcal{I}}[q^4]_{i0}^{33} + 2 \,\tilde{\mathcal{I}}[q^4]_{i0}^{42} + 2 \,\tilde{\mathcal{I}}[q^4]_{i0}^{51} \right) = \frac{1}{6M_i^4} \\
\tilde{f}_{13B}^{i} &= 4\left(\,\tilde{\mathcal{I}}[q^4]_{i0}^{33} + 2 \,\tilde{\mathcal{I}}[q^4]_{i0}^{24} + 2 \,\tilde{\mathcal{I}}[q^4]_{i0}^{15} \right) = -\frac{1}{4 M_i^4} \\
\tilde{f}_{14}^{i} &= -8\,\tilde{\mathcal{I}}[q^4]_{i}^{6}=-\frac{1}{60 M_i^4} \\
\tilde{f}_{14A}^{i} &=  - 8 \,\tilde{\mathcal{I}}[q^4]_{i0}^{33} = -\frac{1}{6 M_i^4} \\
\tilde{f}_{15}^{i} &= 8\,\tilde{\mathcal{I}}[q^4]_{i}^{6} = \frac{1}{60 M_i^4} \\
\tilde{f}_{15A}^{i} &= 4 \left( \,\tilde{\mathcal{I}}[q^4]_{i0}^{33} + \tilde{\mathcal{I}}[q^4]_{i0}^{42} \right)   = \frac{1}{9 M_i^4} \\
\tilde{f}_{15B}^{i} &= 4 \left( \,\tilde{\mathcal{I}}[q^4]_{i0}^{33} + \tilde{\mathcal{I}}[q^4]_{i0}^{24} \right) = -\frac{1}{6 M_i^4}\left( \frac{7}{3}- \log\frac{M_i^2}{\mu^2}\right) \\
\tilde{f}_{16}^{i} &= \frac{1}{5}\,\tilde{\mathcal{I}}_{i}^{5} = -\frac{1}{60 M_i^6} \\
\tilde{f}_{16A}^{i} &= \,\tilde{\mathcal{I}}_{i0}^{41} = -\frac{1}{3 M_i^6} \\
\tilde{f}_{16B}^{i} &= \,\tilde{\mathcal{I}}_{i0}^{32} = \frac{1}{M_i^6}\left( \frac{5}{2} - \log \frac{M_i^2}{\mu^2}\right) \\
\tilde{f}_{16C}^{i} &= \,\tilde{\mathcal{I}}_{i0}^{32} = \frac{1}{M_i^6}\left( \frac{5}{2} - \log \frac{M_i^2}{\mu^2}\right) \\
\tilde{f}_{16D}^{i} &= \,\tilde{\mathcal{I}}_{i0}^{23} = -\frac{2}{M_i^6}\left( \frac{3}{2} - \log \frac{M_i^2}{\mu^2}\right) \\
\tilde{f}_{16E}^{i} &= \,\tilde{\mathcal{I}}_{i0}^{23} = -\frac{2}{M_i^6}\left( \frac{3}{2} - \log \frac{M_i^2}{\mu^2}\right) \\
\tilde{f}_{16F}^{i} &= \,\tilde{\mathcal{I}}_{i0}^{14} = \frac{1}{M_i^6}\left( 1- \log \frac{M_i^2}{\mu^2}\right) \\
\tilde{f}_{17}^{i} &= 6\,\tilde{\mathcal{I}}[q^2]_{i}^{6}  = -\frac{1}{20 M_i^6} \\
\tilde{f}_{17A}^{i} &= 2 \left( \,\tilde{\mathcal{I}}[q^2]_{i0}^{42} + 2\,\tilde{\mathcal{I}}[q^2]_{i0}^{51} \right) = -\frac{1}{4 M_i^6} \\
\tilde{f}_{17B}^{i} &= 6\,\tilde{\mathcal{I}}[q^2]_{i0}^{42} = -\frac{1}{2 M_i^6}  \\
\tilde{f}_{17C}^{i}&= 2 \left(\,\tilde{\mathcal{I}}[q^2]_{i0}^{42} + 2\,\tilde{\mathcal{I}}[q^2]_{i0}^{51} \right)  = -\frac{1}{4 M_i^6} \\
\tilde{f}_{17D}^{i} &= 2 \left( 2 \,\tilde{\mathcal{I}}[q^2]_{i0}^{33} + \,\tilde{\mathcal{I}}[q^2]_{i0}^{42} \right) = \frac{1}{M_i^6}\left( \frac{17}{6} - \log \frac{M_i^2}{\mu^2}\right) \\
\tilde{f}_{17E}^{i} &= 2 \left( \,\tilde{\mathcal{I}}[q^2]_{i0}^{24} + 2\,\tilde{\mathcal{I}}[q^2]_{i0}^{33} \right) = \frac{1}{M_i^6} \\
\tilde{f}_{17F}^{i} &=  2 \left( 2\,\tilde{\mathcal{I}}[q^2]_{i0}^{15} + \,\tilde{\mathcal{I}}[q^2]_{i0}^{24} \right) = -\frac{1}{2 M_i^6} \\
\tilde{f}_{17G}^{i} &=  2 \left( 2\,\tilde{\mathcal{I}}[q^2]_{i0}^{15} + \,\tilde{\mathcal{I}}[q^2]_{i0}^{24} \right) = -\frac{1}{2 M_i^6} \\
\tilde{f}_{17H}^{i} &= 6\,\tilde{\mathcal{I}}[q^2]_{i0}^{24}  = -\frac{3}{M_i^6}\left( 2 - \log \frac{M_i^2}{\mu^2}\right) \\
\tilde{f}_{17I}^{i} &=  2 \left( \,\tilde{\mathcal{I}}[q^2]_{i0}^{24} + 2\,\tilde{\mathcal{I}}[q^2]_{i0}^{33} \right) = \frac{1}{ M_i^6} \\
\tilde{f}_{17J}^{i} &= 2 \left( \,\tilde{\mathcal{I}}[q^2]_{i0}^{42} + 2 \,\tilde{\mathcal{I}}[q^2]_{i0}^{33} \right) =\frac{1}{M_i^6}\left( \frac{17}{6} - \log \frac{M_i^2}{\mu^2}\right) \\
\tilde{f}_{18}^{i} &= 4\,\tilde{\mathcal{I}}[q^2]_{i}^{6}=  -\frac{1}{30 M_i^6} \\
\tilde{f}_{18A}^{i} &= 4\left(  \,\tilde{\mathcal{I}}[q^2]_{i0}^{42} + \,\tilde{\mathcal{I}}[q^2]_{i0}^{51} \right)  = -\frac{5}{12 M_i^6}   \\
\tilde{f}_{18B}^{i} &= 2 \left( 2\,\tilde{\mathcal{I}}[q^2]_{i0}^{33} + \,\tilde{\mathcal{I}}[q^2]_{i0}^{42} + \tilde{\mathcal{I}}[q^2]_{i0}^{24} \right) = \frac{5}{6 M_i^6} \\
\tilde{f}_{18C}^{i} &= 8 \,\tilde{\mathcal{I}}[q^2]_{i0}^{33} =  \frac{2}{M_i^6}\left( 3- \log \frac{M_i^2}{\mu^2}\right) \\
\tilde{f}_{18D}^{i} &=4 \left( \,\tilde{\mathcal{I}}[q^2]_{i0}^{15} + \,\tilde{\mathcal{I}}[q^2]_{i0}^{24} \right) = -\frac{1}{M_i^6}\left( \frac{5}{2} - \log \frac{M_i^2}{\mu^2}\right)  \\
\tilde{f}_{18E}^{i} &= \,\tilde{\mathcal{I}}[q^2]_{i0}^{24} + 2\,\tilde{\mathcal{I}}[q^2]_{i0}^{33} + \,\tilde{\mathcal{I}}[q^2]_{i0}^{42}  = \frac{5}{12 M_i^6} \\  
\tilde{f}_{19}^{i} &= \frac{1}{6} \,\tilde{\mathcal{I}}_{i}^{6} = \frac{1}{120 M_i^8} \\
\tilde{f}_{19A}^{i} &= \,\tilde{\mathcal{I}}_{i0}^{51} = \frac{1}{4 M_i^8} \\
\tilde{f}_{19B}^{i} &= \,\tilde{\mathcal{I}}_{i0}^{42} = -\frac{1}{M_i^8}\left( \frac{17}{6} - \log \frac{M_i^2}{\mu^2}\right) \\
\tilde{f}_{19C}^{i} &= \,\tilde{\mathcal{I}}_{i0}^{42} = -\frac{1}{M_i^8}\left( \frac{17}{6} - \log \frac{M_i^2}{\mu^2}\right) \\
\tilde{f}_{19D}^{i} &= \,\tilde{\mathcal{I}}_{i0}^{33} = \frac{3}{M_i^8}\left( \frac{11}{6} - \log \frac{M_i^2}{\mu^2}\right) \\
\tilde{f}_{19E}^{i} &= \frac{1}{2}\,\tilde{\mathcal{I}}_{i0}^{42} = -\frac{1}{2 M_i^8}\left( \frac{17}{6} - \log \frac{M_i^2}{\mu^2}\right) \\
\tilde{f}_{19F}^{i} &= \,\tilde{\mathcal{I}}_{i0}^{33} = \frac{3}{M_i^8}\left( \frac{11}{6} - \log \frac{M_i^2}{\mu^2}\right) \\
\tilde{f}_{19G}^{i} &= \,\tilde{\mathcal{I}}_{i0}^{24} = -\frac{3}{M_i^8}\left( \frac{4}{3} - \log \frac{M_i^2}{\mu^2}\right) \\
\tilde{f}_{19H}^{ijk} &= \frac{1}{3}\,\tilde{\mathcal{I}}_{i0}^{33} =  \frac{1}{M_i^8}\left( \frac{11}{6} - \log \frac{M_i^2}{\mu^2}\right) \\
\tilde{f}_{19I}^{i} &= \,\tilde{\mathcal{I}}_{i0}^{24} = -\frac{3}{M_i^8}\left( \frac{4}{3} - \log \frac{M_i^2}{\mu^2}\right) \\
\tilde{f}_{19J}^{i} &= \frac{1}{2}\,\tilde{\mathcal{I}}_{i0}^{24} = -\frac{3}{2 M_i^8}\left( \frac{4}{3} - \log \frac{M_i^2}{\mu^2}\right) \\
\tilde{f}_{19K}^{i} &= \,\tilde{\mathcal{I}}_{i0}^{15} = \frac{1}{M_i^8}\left( 1 - \log \frac{M_i^2}{\mu^2}\right) 
\end{align*}
\end{scriptsize}
\end{multicols}

\providecommand{\href}[2]{#2}\begingroup\raggedright

\end{document}